\documentclass[twocolumn,showpacs]{revtex4} 
\usepackage{times,xspace} 
\usepackage{amsbsy,amssymb,amsmath,bm} 
\usepackage{graphicx,color,epsfig,rotate} 
\usepackage{fancyhdr} 
 
\def\bbbc{{\mathchoice {\setbox0=\hbox{$\displaystyle\rm C$}\hbox{\hbox 
to0pt{\kern0.4\wd0\vrule height0.9\ht0\hss}\box0}} 
{\setbox0=\hbox{$\textstyle\rm C$}\hbox{\hbox 
to0pt{\kern0.4\wd0\vrule height0.9\ht0\hss}\box0}} 
{\setbox0=\hbox{$\scriptstyle\rm C$}\hbox{\hbox 
to0pt{\kern0.4\wd0\vrule height0.9\ht0\hss}\box0}} 
{\setbox0=\hbox{$\scriptscriptstyle\rm C$}\hbox{\hbox 
to0pt{\kern0.4\wd0\vrule height0.9\ht0\hss}\box0}}}}

\newcommand{\ignore}[1]{} 
\newcommand{\mComment}[1]{} 
\newcommand{\gComment}[1]{} 
\newcommand{\jComment}[1]{} 
\newcommand{\rComment}[1]{} 
\newcommand{\lComment}[1]{} 
 
\renewcommand{\mComment}[1]{\textcolor{blue}{Bruce: #1}} 
\renewcommand{\gComment}[1]{\textcolor{red}{Zohar: #1}} 
\renewcommand{\jComment}[1]{\textcolor{green}{Cristian: #1}} 
 
\begin{document} 
 
\title{High--dimensional fractionalization and spinon deconfinement in  
pyrochlore antiferromagnets}  
 
\author{Z. Nussinov$^{1}$, C. D. Batista$^{2}$, B. Normand$^{3}$, and  
S. A. Trugman$^{2}$} 
 
\affiliation{$^1$Washington University, Dept. of Physics -- Compton  
Hall, 1 Brookings Drive, St. Louis, MO 63130, USA} 
 
\affiliation{$^2$Theoretical Division, Los Alamos National Laboratory,  
Los Alamos, NM 87545, USA} 
 
\affiliation{$^3$Centro At\'{o}mico Bariloche and Instituto Balseiro,  
Comisi\'{o}n Nacional de Energ\'{\i }a At\'{o}mica, 8400 Bariloche, Argentina} 
 
\date{Received \today } 
 
\begin{abstract} 

The ground states of Klein type spin models on the pyrochlore
and checkerboard lattice
are spanned by the set of singlet dimer 
coverings, and thus possess an extensive ground--state degeneracy. 
Among the many exotic consequences is the presence of deconfined  
fractional excitations (spinons) which propagate through the entire system.  
While a realistic electronic model on the pyrochlore lattice is close to  
the Klein point, this point is in fact inherently unstable because any  
perturbation $\epsilon$ restores spinon confinement at $T = 0$. We  
demonstrate that deconfinement is recovered in the finite--temperature  
region $\epsilon \ll T \ll J$, where the deconfined phase can be  
characterized as a dilute Coulomb gas of thermally excited spinons.  
We investigate the zero--temperature phase diagram away from the Klein  
point by means of a variational approach based on the singlet dimer  
coverings of the pyrochlore lattices and taking into account their  
non--orthogonality. We find that in these systems,
nearest neighbor exchange interactions do not lead to 
Rokhsar-Kivelson type processes.
 
\end{abstract} 
 
 
\maketitle 
 
\section{Introduction}  
 
Discovering new phases of matter is a primary objective of physics. The  
fractionalized spin liquid in two spatial dimensions ($d = 2$) has  
provided a popular candidate framework for models describing the exotic  
properties observed in many strongly correlated electronic materials,  
including frustrated quantum magnets and high--$T_c$ superconductors  
\cite{Anderson87}. The spin--liquid state is characterized by spinon  
excitations carrying unit charge under a compact U(1) gauge field.  
However, Polyakov has argued \cite{Polyakov77} that a pure compact  
U(1) gauge theory is always confining at zero temperature for $d = 2$,  
confinement between test particles with opposite charges being produced  
by the proliferation of instanton tunneling events.  
 
By contrast, the case for confinement by instanton proliferation in spin  
systems is rather more involved \cite{Herbut03, Herm}, and the situation  
at a critical point offers additional possibilities. These considerations 
underlie a recent discussion of the deconfined quantum critical point  
\cite{senthil04} as the foundation for the emergence of spin--liquid  
phases with fractional excitations. However, while this set of elegant  
ideas appears most plausible, it is not yet clear from any analytical  
or numerical calculations that there exists a microscopic Hamiltonian,  
and particularly one with only simple electronic interactions,  
exemplifying this class of physical phenomena. 
Quasiparticle fractionalization has been demonstrated rigorously in one  
dimension (1d) and in the quantum Hall effect. For frustrated quantum  
spin systems, recent progress was made in the context of a two--dimensional  
model with high degeneracy in the ground--state manifold which exhibits  
(partial) dimensional reduction \cite{BN} and consequent spinon  
deconfinement \cite{rbt}. We will expand upon and systematize these  
ideas to provide a further demonstration of a class of physical systems  
displaying fractional excitations.  
 
To this end, we begin with the natural question of whether a dilute gas of  
deconfined spinons can be stabilized in dimensions $d > 1$ in a finite region  
of the phase diagram. This latter requirement is crucial for observing  
deconfined spinon excitations in real systems. We will provide an  
affirmative answer by considering a highly frustrated $S = 1/2$ model  
on the $d = 2$ (checkerboard) and $d = 3$ pyrochlore lattices. We  
demonstrate that for $T = 0$ the spinon excitations are deconfined at  
the Klein critical point, whose nature we will explain in detail. As  
expected from above, this zero--temperature deconfined phase is unstable  
under any realistic perturbation characterized by an energy scale  
$\epsilon$. We argue that in the regime $\epsilon \ll J$, where $J$ is  
the characteristic energy scale of the model, a dilute Coulomb gas of  
spinons is stabilized in the temperature range $\epsilon \ll T \ll J$.  
This result raises the possibility of observing spinons in real  
higher--dimensional systems. 

\begin{figure}[t!]
\vspace*{-3.3cm}
\hspace*{1.5cm}
\includegraphics[angle=90,width=7.0cm]{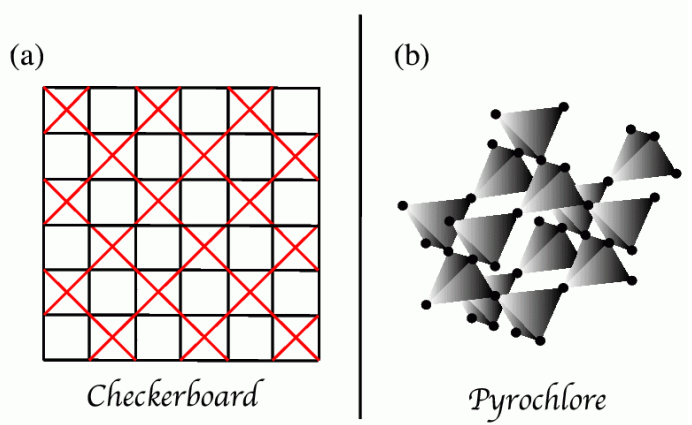}
\vspace{-1.3cm}
\caption{(Color online.) (a) Checkerboard and (b) pyrochlore lattices.}
\label{plattice}
\end{figure}

The checkerboard and pyrochlore lattices (Fig.~\ref{plattice}) which  
form the focus of our analysis have long been known to possess a geometry  
which is amenable to very large ground--state degeneracies. The pyrochlore  
lattice is composed of corner--sharing tetrahedra, arranged in a cubic (fcc)  
structure, such that each vertex is common to two 4--site units (see  
panel (b) of Fig.~\ref{plattice}). This geometry is rather common in  
transition--metal and rare--earth oxide systems, specifically in the  
pyrochlore (the origin of the name) and spinel structures. The pyrochlore  
systems \cite{rramirez} have chemical formula A$_2$B$_2$O$_7$, where both  
A and B may be taken from a wide range of metal ions, particularly  
rare--earths, and both occur on a pyrochlore lattice. Spinel systems  
\cite{rtakagi} have chemical formula AB$_2$O$_4$, where it is more  
typical for the A and B ions to be transition metals and only the B  
sites form a pyrochlore structure. Individual compounds in these classes  
are usually strongly insulating and of fixed ionic valences, and the  
presence of non--magnetic ions on one of the sublattices (the A sites  
for the spinel) leads to a pyrochlore lattice of interacting spins on  
the other. It was noted at a very early stage \cite{fowler,Pauling}  
that the connectivity of such a structure is not dissimilar to the  
arrangement found in water ice, a point which we will both explain  
and exploit extensively below.  
 
The checkerboard lattice can be considered as a 2d version of the 
pyrochlore structure, in the sense that it also is composed of 
corner--sharing tetrahedral units, these being represented by the 
crossed plaquettes (the terminology we adopt henceforth) in panel (a)  
of Fig.~\ref{plattice}. It is the relatively low coordination number 
and tetrahedral connectivity of the checkerboard and pyrochlore lattices 
which permits a proliferation of ground states. Some of the essential 
considerations are presented for classical pyrochlore antiferromagnets 
in Ref.~\cite{frustration} and for quantum pyrochlore antiferromagnets 
in Ref.~\cite{Canals}. While a significant body of recent work has led 
to a better understanding of ice--like phases in pyrochlore ferromagnets 
\cite{gingras}, the situation for both classical and quantum 
antiferromagnets remains far from satisfactory, to the extent that 
certain results conflict \cite{conflict}. Here we provide a further 
contribution to the body of knowledge concerning quantum pyrochlore 
antiferromagnets. 
 
We will demonstrate further that the highly unconventional properties  
listed above arise within simple, nearest--neighbor, bilinear and  
biquadratic spin models on the 2d and 3d pyrochlore lattices. These models  
are derived directly from half--filled, nearest--neighbor Hubbard models,  
{\it i.e.}~from models with Coulomb interactions of the shortest range,  
and are thus fully realistic for electronic systems. The key features  
giving rise to exotic behavior are (i) an exponential number of degenerate  
ground states; (ii) exact critical behavior with an emergent divergent  
correlation length, accompanied by the absence of long--ranged order;  
(iii) fractionalization of excited spin states into spinons which  
propagate freely in space, implying a separation of spin and charge  
upon doping these systems (we will show that the emergent fractionalized  
excitations have well--defined quantum numbers which may be described as  
residing at the ends of strings obeying simple rules); (iv) an effective  
dimensional reduction occurring precisely at the critical point; (v)  
topological order, in a sense which we will define explicitly below.  
All of these features lead in combination to a distinct type of  
critical point whose static and dynamic properties we will characterize  
in detail, illustrating in addition their connection to a multitude of  
ideas and concepts emerging in the recent literature in the context of  
deconfinement and criticality. 
 
In Sec.~II we introduce the type of spin model displaying a Klein point,  
and discuss its foundation in the electronic Hamiltonian of real, correlated  
insulating materials. In Sec.~III we begin our analysis of the model at the 
Klein critical point and at zero temperature, focusing on the characteristic 
properties here and illustrating their origin in the extensive ground--state 
degeneracy. In Sec.~IV we remain at the Klein point but consider finite 
temperatures, to derive the effective entropic interaction of spinon 
excitations and thus demonstrate their thermally driven deconfinement. In 
Secs.~V and VI we restore some of the perturbing effects present in the 
realistic Hamiltonian of Sec.~II. Working at zero temperature, in Sec.~V 
we introduce and apply a specifically constructed variational procedure 
based on the (non--orthogonal) dimer coverings of the ground--state manifold, 
to demonstrate the selection of specific valence--bond orderings as the ground 
states of the perturbed system. In Sec.~VI we extend our considerations to 
finite temperatures and present one example of an exactly critical regime. 
These calculations indicate explicitly both the confinement of spinons at 
low temperatures and their liberation at higher temperatures, thus 
reinforcing the results of Secs.~III and IV. Section VII contains a 
discussion and conclusion. 
 
\section{Model} 
 
The spin model upon which we focus is a $S = 1/2$ Hamiltonian on the  
checkerboard and pyrochlore lattices, 
\begin{equation} 
H = J \sum_{\langle ij \rangle, \alpha}  H_{ij}^{\alpha} + K \sum_{\alpha}  
(H_{ij}^{\alpha} H_{kl}^{\alpha} + H_{il}^{\alpha} H_{jk}^{\alpha} +  
H_{ik}^{\alpha} H_{jl}^{\alpha}),  
\label{esh} 
\end{equation} 
where $\langle ij \rangle$ denotes all pairs of sites in the tetrahedron 
$\alpha \equiv ijkl$ and $H_{ij}^{\alpha} = {\vec S}_i^{\alpha} \cdot  
{\vec S}_j^{\alpha}$ is a scalar product of the two spins. On the  
checkerboard lattice $\alpha$ denotes each crossed plaquette, by which is  
meant those in Fig.~\ref{plattice}(a) with cross--coupling interactions, 
and the first term is equivalent to both nearest-- and next--neighbor  
Heisenberg interactions of strength $J$. This model was considered on  
the square lattice (all plaquettes crossed) by two of the present  
authors \cite{rbt}, and we will contrast the two $d = 2$ systems below.  
The Heisenberg interaction on the tetrahedral units (we will also refer  
to the crossed plaquettes of the checkerboard lattice as tetrahedra) of  
the $d = 2$ and $d = 3$ pyrochlore lattices is strongly frustrated, and  
the symmetric four--spin interaction pushes the model to maximal  
frustration at $K_c = 4J/5$ \cite{explain_frustration}.  

A straightforward but important observation is that $H$ may be recast in  
the form 
\begin{equation}  
H = \frac{J_{1}}{2} \sum_{\boxtimes} {\vec S}_{\boxtimes}^{2}  
+ \frac{J_{2}}{4} \sum_{\boxtimes} {\vec S}_{\boxtimes}^{4}, 
\label{symHam} 
\end{equation}  
with ${\vec S}_{\boxtimes}$ the net spin of the tetrahedral unit $\boxtimes$, 
\begin{equation} 
{\vec S}_{\boxtimes} \equiv \sum_{i \in \boxtimes} {\vec S}_{i}, 
\end{equation}  
$J_1 = J - 7K/4$ and $J_2 = K/2$ \cite{simple_identities}.  
This Hamiltonian (\ref{esh}) can be obtained from the Hubbard model on  
the pyrochlore lattice,  
\begin{equation} 
H_{\rm Hubb} = - t \sum_{\langle ij \rangle, \sigma}  c_{i\sigma}^{\dag}  
c_{j\sigma} + U \sum_{i} n_{i\uparrow} n_{i\downarrow},  
\label{eeh} 
\end{equation} 
where $n_{i\sigma} = c_{i\sigma}^{\dag} c_{i\sigma}$ is the number operator  
for electrons of spin $\sigma$. This is perhaps the simplest possible  
description of interacting electrons with only nearest--neighbor hopping  
$t$ and a local Coulomb repulsion $U$. Systems which are half--filled and  
have strong on--site interactions are localized insulators where, with  
the exception of virtual processes, the kinetic energy gain from hopping  
is sacrificed to avoid the Coulomb cost of double occupancy. The effective  
Hamiltonian obtained from the virtual hopping processes at lowest (second)    
order in the small parameter $t/U$ is a nearest--neighbor Heisenberg  
model of the type contained in the first term of Eq.~(\ref{symHam}), with  
$J_1 = 4 t^2 / U$ and electronic spin ${\vec S}_i^a = c_{i\beta}^{\dag}  
\sigma_{\beta\gamma}^a c_{i\gamma}$ ($a = x,y,z$).  
 
A fourth--order strong--coupling expansion of Eq.~(\ref{eeh}) returns  
interaction terms bilinear in spins on the same and on different tetrahedra 
and biquadratic in spins on the same tetrahedron \cite{note0}. As shown in  
detailed studies performed for plaquettes of the square lattice in the  
context of planar cuprates \cite{rmhr}, and verified by fits to magnon  
dispersion relations for the same systems \cite{rkk}, the latter  
interactions are stronger than the former by approximately one order  
of magnitude. Specifically, a far larger number of intra--tetrahedron  
processes, whose sum has the symmetric form contained in the second  
term of Eq.~(\ref{esh}), contributes to the effective fourth--order  
Hamiltonian than do processes of the next--neighbor Heisenberg type.  
Thus by retaining only the intra--tetrahedron processes to fourth order  
in $t$ one obtains the spin Hamiltonian of Eq.~(\ref{symHam}) with 
\cite{rcgb}
\begin{eqnarray} 
J_{1} & = & \frac{4t^{2}}{U} - \frac{160 t^{4}}{U^{3}} + {\cal{O}} \left(  
\frac{t^{6}}{U^{5}} \right), \nonumber \\ J_{2} & = & \frac{40 t^{4}}{U^{3}}  
+ {\cal{O}} \left( \frac{t^{6}}{U^{5}} \right). 
\label{j12}
\end{eqnarray} 
The general, SU(2)--invariant spin Hamiltonian arising from a realistic  
electronic model may thus be considered as an intra--plaquette interacting  
system of exactly the form given in Eq.~(\ref{symHam}), with only very weak  
perturbations from inter--tetrahedron terms. To quantify this statement, in 
the half--filled system Eqs.~(\ref{symHam}) and (\ref{eeh}) are related by
\begin{eqnarray}
{\tilde H}_{\rm Hubb} = H + J_{3} \sum_{\langle \langle i j \rangle \rangle}
\vec{S}_{i} \cdot \vec{S}_{j},
\label{compare}
\end{eqnarray}
where ${\tilde H}_{\rm Hubb}$ is the effective spin Hamiltonian obtained
from a fourth--order strong--coupling expansion of $H_{\rm Hubb}$. 
The sum is over pairs of sites $\langle \langle i, j \rangle \rangle$ 
separated by two bonds and which do not belong to the same tetrahedron. 
The exchange coupling $J_3$ is given by 
\begin{eqnarray}
J_{3} = \frac{4t^{4}}{U^{3}} + {\cal{O}} \left(\frac{t^{6}}{U^{5}}
\right).
\label{j3}
\end{eqnarray}
It is evident from Eqs.~(\ref{j12}) and (\ref{j3}) that for small $t/U$ 
the inter--plaquette corrections ($J_{3}$) originating from the Hubbard 
Hamiltonian are indeed one order of magnitude smaller than 
the intra--plaquette terms of $H$ (\ref{symHam}). A particle-hole transformation 
leaves the spectrum of $H_{\rm Hubb}$ invariant at half-filling. Since the net effect of
this transformation is to change the sign of the hopping, ${\tilde H}_{\rm Hubb}$ 
cannot contain any term of odd order in $t$. 

In Sec.~III we will analyze the properties of Eq.~(\ref{symHam}) at the  
Klein point, $J_1 = - J_2$ \cite{ferrom} or $K = K_c = 4J/5$, where 
the semi--positive  
definite Hamiltonian of Eq.~(\ref{symHam}) attains its minimal value (when 
$S_{\boxtimes} = 0$ or 1 on every tetrahedron). 
The Klein point is quasi-exactly solvable. [Its ground
states may be determined exactly.]
Following the observations  
in the preceding paragraph, in Secs.~V and VI we will expand our treatment 
to include perturbations around the Klein point, which are of two types: 
the stronger, represented by nearest--neighbor Heisenberg terms,  
correspond to deviations of Eq.~(\ref{esh}) from the point $K = K_c$;  
the weaker, represented by next--neighbor Heisenberg terms (and, for the  
checkerboard, cyclic terms on the uncrossed plaquettes), correspond to  
fourth--order inter--plaquette interactions. Specific $S = 1/2$ materials  
can be expected at least in the 3d pyrochlore structure, such as that  
formed by the magnetic ions at both A$^{m+}$ and B$^{n+}$ ($m + n = 7$)  
sites in the stereotypical pyrochlore systems A$_2$B$_2$O$_7$, and on  
the B sites in A$^{m+}$B$_2^{n+}$O$_4$ ($m + 2 n = 8$) spinels. However,  
despite the considerable choice of magnetic and non--magnetic ionic  
species afforded by these two structural classes, to date we are not  
aware of the successful synthesis of any such compounds which maintain  
the full cubic lattice symmetry. 
 
\section{Exact ground states for the $S = 1/2$ system at the Klein point} 
 
Equation (\ref{symHam}) suggests an even simpler way of expressing the  
Hamiltonian,  
\begin{equation} 
H = \sum_{\boxtimes} h_{\boxtimes} 
\label{ho} 
\end{equation} 
with $h_{\boxtimes}$ a quartic polynomial in the total spin of the four sites  
forming a given tetrahedral unit of the pyrochlore lattice, $h_{\boxtimes} =  
\frac{1}{2} J_1 S_{\boxtimes}^{2} + \frac{1}{4} J_2 S_{\boxtimes}^{4}$. For  
clarity we provide a brief review of the possible two-- and four--particle  
states of $S = 1/2$ spins. For a system of four coupled spins, the total  
spin $S_{\boxtimes}$ of any tetrahedron obeys $0 \le S_{\boxtimes} \le 2$. 
The 16--state spin space is decomposed into two singlet states ($S_{\boxtimes} 
= 0$), three triplets $(S_{\boxtimes} = 1)$, and one quintet ($S_{\boxtimes} 
= 2$), 
\begin{equation} 
\frac{1}{2} \otimes \frac{1}{2} \otimes \frac{1}{2} \otimes \frac{1}{2} 
= 0 \oplus 0 \oplus 1 \oplus 1 \oplus 1 \oplus 2. 
\label{Hs} 
\end{equation} 
In this explicit decomposition the right--hand side labels the disjoint 
net spin ($S_{\boxtimes}$) sectors while the left--hand side encodes  
the 16--dimensional space spanned by the direct product of the four $S =  
1/2$ particles. The sum of any two nearest--neighbor spins on the lattice  
$({\vec S}_{i} + {\vec S}_{j})^{2} = S_{\rm pair}(S_{\rm pair} + 1)$ takes  
only the two values $S_{\rm pair} = 0,1$. If any two spins within a given  
tetrahedron are in a singlet state ($S_{\rm pair} = 0$), then the total 
spin of this tetrahedron cannot exceed unity, $S_{\boxtimes} \le 1$.   
For the parameter choice $K = K_c$ ($J_2 = - J_1$), $H$ (\ref{symHam}) 
can be rewritten, up to irrelevant constants, as a Klein Hamiltonian 
\cite{Klein82}, 
\begin{equation} 
H_K =  \frac{12}{5} J \sum_{\boxtimes} {\cal{P}}^{\boxtimes}, 
\label{GK} 
\end{equation} 
where ${\cal{P}}$ is the projection operator onto the subspace of  
net spin $S_{\boxtimes} = 2$.  
 
\begin{figure}[t!]
\vspace*{-0.5cm}
\includegraphics[angle=0,width=9cm]{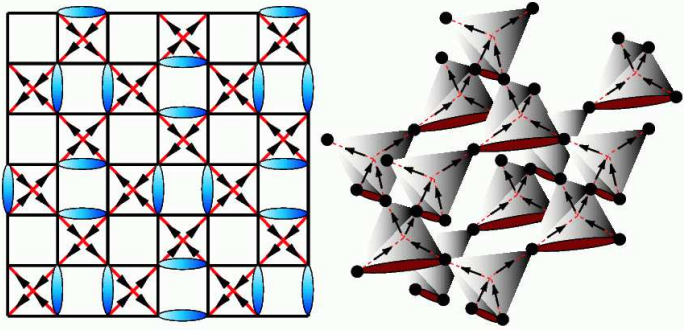}
\vspace{-1.5cm}
\caption{ (Color online.) Highly regular ground states on the checkerboard 
and pyrochlore lattices. The ovals denote singlet dimer states. The arrows 
denote the representations of these dimer states within the six--vertex 
model (see text). On each plaquette (tetrahedron) the dimer connects the 
bases of the two incoming arrows.}
\label{PR}
\bigskip 
\end{figure}

Next we note, following Ref.~\cite{Raman05},  
that because the number of lattice sites is double the number of tetrahedra, 
$N_s = 2N_t$, the set of ground states with one singlet per tetrahedron can  
be mapped onto the set of hard--core dimer coverings of the lattice. Thus  
any dimer covering of the lattice with precisely one dimer per  
tetrahedron is a ground state of $H_K$. Further, any state $| \psi \rangle$  
which is a superposition of dimer coverings each of which has one dimer per  
tetrahedron, 
\begin{equation} 
| \psi \rangle \! = \! \sum_{P} \alpha_{P} \! \prod_{ij \in P} \! |S_{ij}  
\rangle,\;\, {\rm with} \;\, | S_{ij} \rangle = {\textstyle \frac{1} 
{\sqrt{2}}} (| \uparrow \downarrow \rangle - | \downarrow \uparrow \rangle) 
\label{dimer}  
\end{equation} 
(the dimer coverings are labeled by $P$), is also a ground state of $H_K$.  
One of the simplest ground states is afforded by the dimer coverings  
depicted in Fig.~\ref{PR}. While it is clear that any state of the form 
of Eq.~(\ref{dimer}) is a ground state, a proof that all ground states 
are of this form is far less obvious. It may nevertheless be shown 
rigorously \cite{nussinov_June06} that for Klein models on the 
pyrochlore lattices all ground states are indeed of the type specified 
in Eq.~(\ref{dimer}).

As a first qualitative consequence of this result, it was noted in  
Ref.~\cite{rbt} for the analogous point in the square--lattice Hamiltonian  
that satisfying this constraint leads to a significant degeneracy because  
dimer singlets may be rearranged along diagonal lines of the lattice with  
no energy cost. The ground--state entropy scales with the perimeter of the  
system, and pairs of single--spin (spinon) excitations on the diagonal  
lines have no binding energy for any separation. Thus the effective  
dimensional reduction from 2d to 1d leads to the presence of deconfined  
spinons whose propagation is essentially unidimensional. On the pyrochlore  
lattice the open structure of crossed plaquettes creates a less constrained  
system, and the number of dimer configurations in the ground--state manifold  
is strongly enhanced, the still more massive degeneracy implying similarly  
exotic physics in this case. We proceed quantitatively by computing the  
ground--state entropy in Sec.~IIIA. 
 
In addition to the simple coverings shown in Fig.~{\ref{PR}, a far richer  
variety of states exists. The mapping onto the spin--ice problem mentioned  
briefly above provides a useful classification of these ground states. We  
stress that the states with one dimer per tetrahedral unit defined by  
Eq.~(\ref{dimer}) map exactly to the spin--ice problem \cite{gingras,  
anderson}} where, motivated by the structure of H atoms in solid water  
\cite{fowler,Pauling}, two of the sites of any elementary unit (a  
tetrahedron) are associated with an ingoing arrow pointing towards the  
center of the tetrahedron and two sites lie on arrows pointing outwards.  
We label the two sites belonging to a singlet dimer $|S_{ij} \rangle$ in  
a given tetrahedron $\boxtimes$ by two incoming arrows from sites $i$ and 
$j$ to the center of the unit $\boxtimes \equiv ijkl$. In this fashion it 
is clear that the system is mapped to a set of continuous directed lines 
such that each tetrahedral unit has exactly two incoming and two outgoing 
arrows relative to its center. An example of this mapping is illustrated in  
Fig.~\ref{PR}. A longer but ultimately equivalent version of this mapping,  
enabling a study of correlations and degeneracies, was reported in  
Ref.~\cite{Raman05}. The mapping is one--to--one: any spin--ice configuration  
determines a unique singlet--covering state with one dimer per tetrahedral  
unit and vice versa. The conservation of incoming and outgoing arrows in  
the spin--ice representation is the foundation for the divergence--free  
condition we will employ below. Throughout this work we focus on systems  
with periodic boundary conditions (PBCs). A system with open boundary  
conditions (OBCs) would possess a number of additional states not captured 
by the six--vertex mapping, the entropy of which scales with its surface.  
 
\subsection{Ground--State Degeneracy and Entropy} 
 
Because the number of pyrochlore spin--ice states is bounded from below  
\cite{Onsager} by the Pauling limit \cite{Pauling}, the number of ground  
states (or, more precisely, of non--orthogonal dimer coverings) is given by  
\begin{equation} 
N_{g} > (3/2)^{N/2}, 
\label{pauling.} 
\end{equation} 
with $N$ the number of vertices of the pyrochlore lattice and $N_{g}$  
the number of ground--state singlet dimer coverings. We emphasize that  
this degeneracy is exponential in the system volume, yielding an extensive  
entropy with a simple, analytical expression for the lower bound, the Pauling
entropy $S_{P} = {\textstyle \frac{1}{2} N k_{\rm B} \ln {\frac{3}{2}}}$. 
The ``exact'' entropy of the ice problem is known from series expansion  
methods \cite{rnagle}, which yield the numerical result 
\begin{eqnarray}
S_g = (0.20501\dots) Nk_{\rm B} > S_P = (0.20273\dots) N k_{\rm B}.
\end{eqnarray}

Similarly, for the checkerboard lattice the system is reduced to the exact  
spin--ice model \cite{Baxter} on the square lattice formed by the centers  
of the crossed plaquettes, leading to an exact singlet dimer covering  
degeneracy \cite{Baxter,Lieb} of  
\begin{equation} 
N_{g} = (4/3)^{3N/4}  
\label{elliot.} 
\end{equation} 
with $N$ the number of vertices of the checkerboard lattice, whence  
$S_g = {\textstyle \frac{3}{4}} N k_{\rm B} \ln {\frac {4}{3}}$. In this  
derivation we have assumed that the singlet dimer coverings, although  
non--orthogonal, are linearly independent \cite{note2}. There are  
additional ``defect'' states in the ground--state manifold, in which  
two singlet dimers may occupy a single tetrahedron, or unpaired spins  
may be present, without altering the total energy \cite{rbtfig}. However,  
because the entropy of these states scales with the system surface they 
constitute a set of measure zero in comparison with the extensive set of 
dimer coverings, and will be neglected in what follows.  
 
\begin{figure}[t!]
\vspace*{-1.0cm}
\hspace*{-0.6cm}
\includegraphics[angle=90,width=7cm]{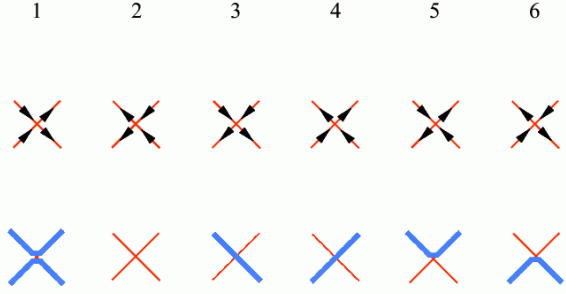}
\vspace{-2.7cm}
\caption{ (Color online.) Standard representation of six--vertex states 
in terms of lines \cite{Baxter}. Every line is composed of links whose 
arrows point to the right in the vertex representation.}
\label{rules}
\end{figure} 
 
\subsection{Geometry of ground states}  
 
To examine the structure of general ground states of the model (\ref{esh})  
at the Klein point, we invoke the line representation of the six--vertex  
model \cite{Baxter} depicted in Fig.~\ref{rules}. In this convention, 
every line is composed of links whose arrows point to the right, and there  
is clear one--to--one mapping between the six--vertex and the line  
representations. An essential property of the lines is their ``chiral''  
nature: they always represent a motion only to the right (Fig.~\ref{rules}).  
We will show in detail below (Fig.~\ref{lightcone}) that this chirality is 
not merely an artificial feature of the line representation, but encodes 
the physical restrictions on the possible spinon paths in the background 
of the allowed dimer states.
 
\begin{figure}[t!]
\vspace*{-0.5cm}
\hspace*{0.0cm}
\includegraphics[angle=90,width=9.0cm]{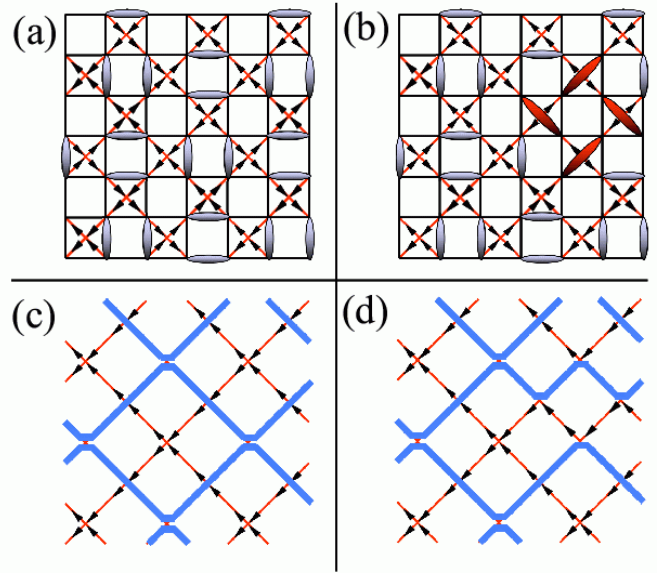}
\vspace*{-1.5cm}
\caption{ (Color online.) Left: representation of six--vertex states in a 
highly regular dimer configuration (a) in terms of lines (c). Joining the 
vertex coverings according to the standard prescription (see text) 
\cite{Baxter} allows one to label all permitted ground states. Right: 
an elementary local spin interaction process, leading to a new dimer 
configuration (b), corresponds in the line representation (d) to 
flipping one of the line corners.}
\label{fig:pr1}
\end{figure}

Finding the elementary process connecting a given dimer covering with  
other ground states is straightforward in the line representation.  
Fig.~\ref{fig:pr1}(c) shows the line representation of the ground state  
depicted in Fig.~\ref{PR}. The simplest process in the dimer basis  
corresponds in the line representation to flipping over one of the line  
corners, as depicted in Figs.~\ref{fig:pr1}(c) and (d). Note that there  
is an uncrossed plaquette (in the original lattice) associated with  
each such corner flip. Returning to the dimer representation, such a  
process corresponds to a cyclic rotation of the four dimers connected  
to the uncrossed plaquette [Fig.~\ref{fig:pr1}(b)]. A similar process  
around a hexagon whose edges belong to six different tetrahedra is  
obtained for the pyrochlore lattice. For clarity of explanation we  
comment that in Figs.~\ref{fig:pr1}(c) and (d), and later in this  
section, the six--vertex (thin lines with arrows) and line representations  
(thick) are superposed to aid in illustrating their equivalence, but either  
alone is a complete representation of the system in terms of independent  
sets of quantum numbers.  

\begin{figure}[t!]
\vspace*{-0.5cm}
\hspace*{-0.5cm}
\includegraphics[angle=90,width=9cm]{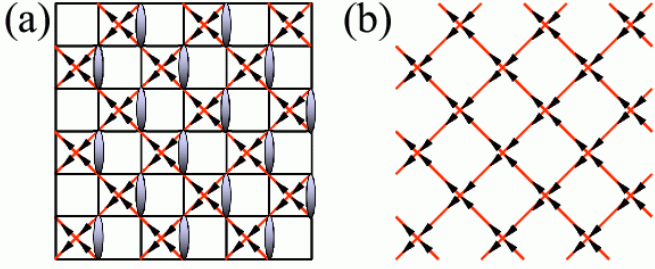}
\vspace*{-5.5cm}
\caption{ (Color online.) Uniform dimer covering or ``ferroelectric'' 
state \cite{Baxter}, corresponding to vacuum of lines. The opposing 
``ferroelectric'' polarization obtained by spatial inversion corresponds 
to the state which is full of lines.}
\label{fig:vac}
\end{figure}

The mapping to the line representation shows that the low--energy sector  
of a Hamiltonian at the Klein point can be mapped into a string model. It  
is clear that the total number of lines is a conserved quantity under any  
local physical process, where by ``local'' we mean that the process changes  
only a finite number of (spin) degrees of freedom. The dimer coverings may  
therefore be classified according to this line quantum number. As an example,  
in Fig.~\ref{fig:vac} we show that the uniform or ``ferroelectric''  
\cite{Baxter} dimer covering corresponds to the vacuum of lines. Number  
conservation implies that this state cannot be connected with any other  
ground state by a local process. The other ``ferroelectric'' dimer state  
(of opposite polarization) which is obtained by spatial inversion  
corresponds in the line representation to the state which is full of  
lines, and by the same argument is not connected by a local process  
to any other states in the ground manifold. Spatial inversion corresponds  
in the line representation to the line--antiline transformation (line  
conjugation).  
 
Not unexpectedly, the exact solution of the six--vertex model \cite{Baxter}  
reveals that the ground states of greatest importance (by which is meant  
those becoming increasingly sharp in the thermodynamic limit) are those  
containing, on a checkerboard lattice of size $L \times L$ , precisely  
$L/2$ lines. This sector has the highest statistical (entropic) weight by  
virtue of the many ground states it possesses. Analogously, we will see  
in Sec.~V that when the system is perturbed away from the Klein point at  
$T = 0$ the state stabilized by the perturbation is generally one with  
$L/2$ lines \cite{note3}. 
 
\subsection{Critical correlations} 
 
To expand upon the previous statements, we stress that all of the dimer  
coverings in the ground--state manifold are individually eigenstates  
of the Hamiltonian. Thus there are no quantum fluctuations at the Klein  
point, and a zero--temperature transition through this point, for example  
as a function of $K/J$, is a discontinuous, first--order quantum phase  
transition \cite{rbt}. The Klein point is a distinct type of critical  
point which is essentially classical in nature with critical thermal  
fluctuations at all temperatures $T > 0$. 
 
It has long been known \cite{Baxter} that the square--lattice ice model  
formed by the centers of the crossed plaquettes in the checkerboard  
lattice exhibits critical power--law correlations, which have been 
investigated extensively in Refs.~\cite{dipole}. Elementary entropic  
arguments combined with the ice condition lead to a simple dipolar analogy  
suggesting that the dimer correlations in $d$ dimensions exhibit an  
$|r|^{-d}$ decay, where ${\bf r}$ is the separation between dimers  
\cite{dipole}. Specifically, denoting by ${\bf P} = (\pm 1, 0,0)$ the  
first two six--vertex configurations of Fig.~\ref{rules}, and similarly  
by ${\bf P} = (0, \pm 1, 0)$ and ${\bf P} = (0,0, \pm 1)$ the other two  
pairs, the line representation applied for the pyrochlore lattice yields  
\cite{dipole}  
\begin{equation} 
\langle P_{i}(0) P_{j}(r) \rangle = \frac{A}{r^3}[\delta_{ij} -  
 3 \hat{r}_{i} \hat{r}_{j}] 
\label{dipole_eqn}
\end{equation} 
for large separation $|{\bf r}|$, with $i,j = 1,2,3$ denoting the  
spatial components of ${\bf P}$ and $\hat{\bf r}$ the unit vector in  
the direction of ${\bf r}$; $A$ is a constant. The correlation function 
for the checkerboard lattice has a similar $|r|^{-2}$ decay. When 
interpreted as flux lines, the six--vertex arrows (Fig.~\ref{rules}) 
adhere at every vertex to a strict condition of no divergence. This 
correspondence, along with the entropic arguments, underlies the dipolar 
correlations. The zero--flux condition at every vertex leads to the 
long--ranged correlations (in fact with infinite correlation length) 
of the dimers. 
 
The ground--state manifold at $K = K_{c}$ corresponds to the 
high--temperature limit of the ice model, in which every six--vertex 
configuration is realized with equal amplitude. As such \cite{Baxter}, 
the system is in the ``disordered'' critical phase described earlier 
where, in particular, the expectation value of any local six--vertex 
order parameter vanishes \cite{Baxter}. On the checkerboard (pyrochlore) 
lattice the different sets of states in the manifold remain characterized 
by the net number of vertical lines on each horizontal row (plane)  
of lattice sites (or conversely): this number is conserved and corresponds  
to a topologically invariant ``flux'' of  lines. Because lines of arrows 
cannot terminate in the six--vertex representation the flux cannot change 
from one plane to the next in the pyrochlore \cite{Baxter}. We remark 
for clarity that for the checkerboard lattice the ``horizontal'' and 
``vertical'' directions specified above for the dual lattice are  
rotated by 45$^0$ relative to the horizontal and vertical directions of  
the original lattice. Although the system does not possess local order 
it displays this elementary topological order characterized by the 
topologically invariant flux \cite{Wenbook}. The model discussed here 
forms a high--dimensional realization of Wen's string nets \cite{Wenbook}. 
On the checkerboard lattice, conservation of the line quantum number means 
that if $n_{Y}$ is the number of vertical lines intercepting a horizontal 
span of fixed length at position $Y$ on the $y$ axis, the non--local 
correlation function for two strings at $y = Y_{1}$ and $y = Y_{2}$, 
which is given by  
\begin{equation} 
\overline{G}(Y_{1}, Y_{2}) \equiv  
\langle e^{i \phi(n_{Y_{1}} - n_{Y_{2}})}\rangle 
\label{topoeq} 
\end{equation} 
with $\phi$ arbitrary, is maximal, $\overline{G}(Y_{1}, Y_{2}) = 1$. 
Because in general $|G| \le 1$, for unrestricted values of $n_{Y_{1}}$ 
and $n_{Y_{2}}$, here one has simply $n_{Y_{1}} = n_{Y_{2}}$. Similar 
results apply for the pyrochlore lattice, where $n_{Y_{i}}$ denotes the 
number of lines intersecting the plane $Y = Y_{i}$. Equation (\ref{topoeq}) 
presents a high--dimensional analog of the well--known fixed value of string 
correlation functions in $S = 1$ spin chains at the AKLT point \cite{AKLT, KT}.
At large separations these uniform non--local correlations dwarf the local 
dimer--dimer correlations, which retain an algebraic decay (as $r^{-d}$). 
As a direct consequence (below), the system exhibits in general a 
long--ranged dimer order away from the Klein critical point.  
 
\subsection{Low--energy excitations at the Klein point: spinons} 
 
\begin{figure}[t!]
\vspace*{-0.3cm}
\hspace*{-0.7cm}
\includegraphics[angle=90,width=9cm]{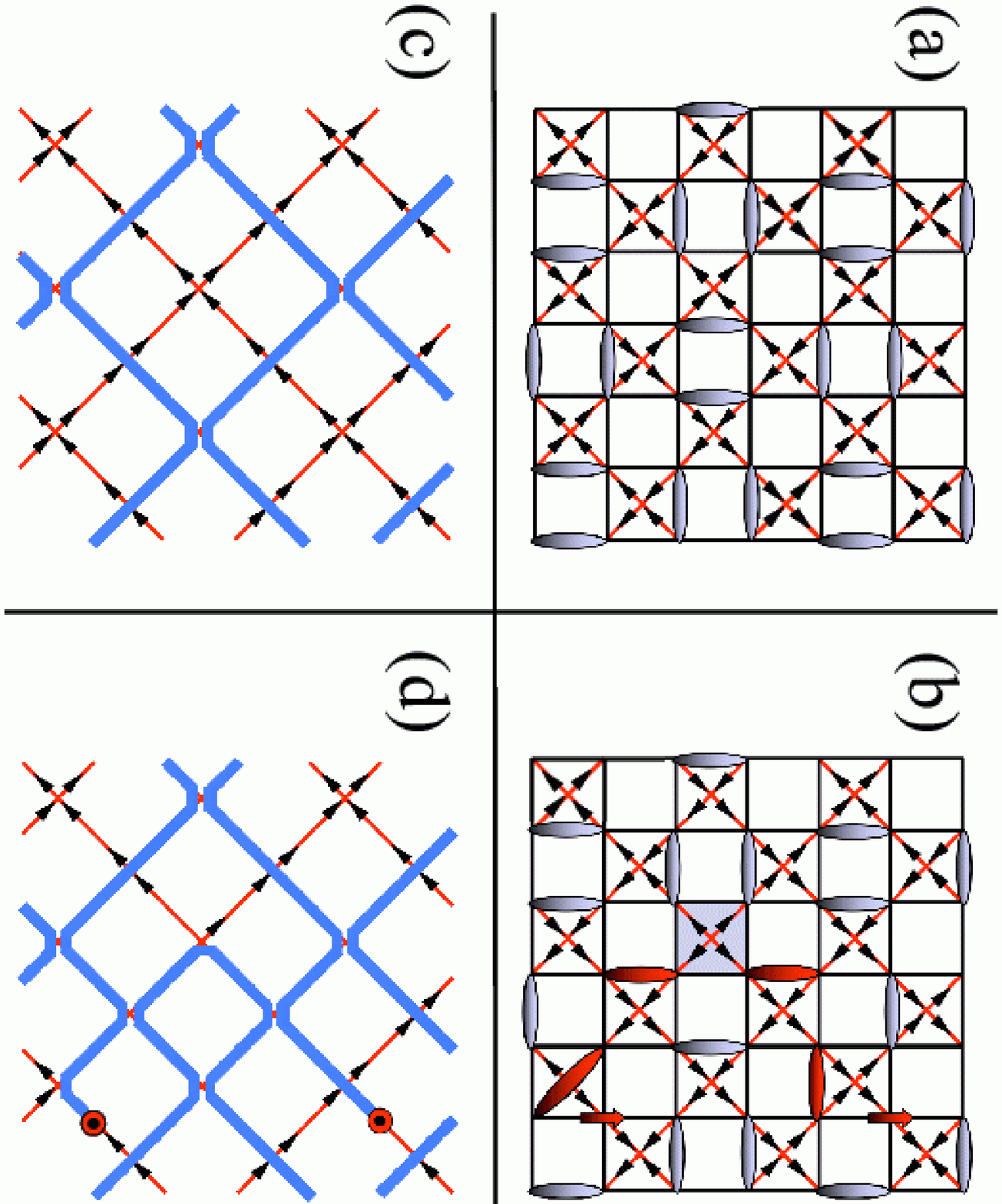}
\vspace{-1.3cm}
\caption{Example of spinon propagation. The potential energy does not 
depend on the length of the string connecting the pair of spinons, which 
consequently are in a deconfined phase. In the line representation the 
position of each spinon is indicated with a small dotted circle.
The shaded square corresponds to a plaquette without singlets that we denote
as a ``defect''.}
\label{spinon}
\end{figure}

The energy cost of exciting one of  
the singlets of a given dimer configuration is proportional to $J$; as  
depicted in Fig.~\ref{spinon}, this energy is not altered by changes in  
the separation of the two $S = 1/2$ objects forming the triplet state,  
which may thus be regarded as two deconfined spinon excitations.  
Single--spinon propagation can be described as a constrained random walk  
because the allowed paths may not increase the number of tetrahedra without  
a singlet dimer, but in contrast to the situation in Ref.~\cite{rbt} the  
paths are $d$--dimensional rather than being constrained to a 
one--dimensional path. 
 
In this context it is instructive to consider the dimensionality of the  
allowed spinon paths for the checkerboard and pyrochlore Klein models, 
to examine its relation to the dimensionality $d_{g}$ of the symmetry  
group emerging in the ground state sector \cite{BN}. Here $d_{g}$ is  
the dimensionality of the minimal non--empty set of spins influenced 
by these symmetry operations. Assuming that the ground--state sector  
does not transform trivially under these symmetry operations, the  
entropy $S_g$ scales according to $S_g \sim N^{d_s/d}$, where $d_s  
= d - d_g$. The Klein model on the pyrochlore lattice exhibits  
a zero--dimensional $(d_{g} = 0$) symmetry, meaning that there exist  
local operations which link different ground states ({\it e.g.}~the  
operation depicted in Fig.~\ref{fig:pr1}).  
 
The presence of these symmetry operators enables spinon excitations to  
propagate with no energy cost in $d_{s}$ dimensions. Thus spinons on  
the pyrochlore and checkerboard lattices propagate freely in regions whose  
size scales with the volume of the entire lattice, $d_s = d$. For the 
square--lattice model of Ref.~\cite{rbt}, $d = 2$ while $d_{g} = 1$, as  
a consequence of which the spinons can propagate only in $d_{s} =  
d - d_{g} = 1$ dimensional regions (lines). This analysis provides a  
more explicit definition of the terms ``complete'' (pyrochlore and  
checkerboard) and ``partial dimensional reduction'' (square lattice),  
of their origin in terms of system entropy (scaling respectively with  
the volume and with the perimeter), and of their consequences for  
spinon dynamics.  
 
As noted earlier, perturbation--theoretic calculations, including those  
illustrating the effects of spinons, are somewhat involved in the dimer  
basis due to the non--orthogonality of the dimer states. In the remainder  
of this section we comment only briefly and qualitatively on possible  
spinon exchange paths connecting different dimer states of the low--energy  
sector; a detailed discussion of one systematic approach to quantitative  
calculations in the dimer basis is deferred to Sec.~V.  
 
\begin{figure}[t!]
\vspace*{-0.3cm}
\hspace*{-0.9cm}
\includegraphics[angle=90,width=9cm]{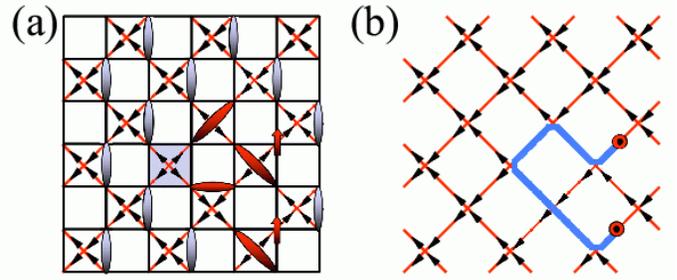}
\vspace{-4.5cm}
\caption{ (Color online.) Spinon propagation in the ``ferroelectric'' or 
uniform state depicted in Fig.~\ref{fig:vac}. Disruption of the dimer 
background dictates that the line or string created between the propagating 
spinons can move only in the direction indicated by the black arrows.
This implies that the spinons are restricted to move within a cone
defined by the horizontal and vertical directions. In the line 
representation, the position of each spinon is indicated with a 
small dotted circle.}
\label{lightcone}
\end{figure}

While we have used the words ``free propagation'' of spinons to indicate  
that their motion occurs in all spatial dimensions, spinon dynamics on  
the pyrochlore and checkerboard lattices at the Klein point remain  
constrained by the effective ice rules obeyed by the dimer configurations  
in the ground--state manifold. These rules, upon which we remark further  
below, become most evident by considering the line representation of the  
six--vertex model (Fig.~\ref{rules}). In the presence of a single dimer  
excitation, the allowed lines specify the possible spinon motion. Against  
the background of the uniform ``ferroelectric'' state (corresponding to  
the line vacuum, Fig.~\ref{fig:vac}), where all horizontal and vertical  
arrows point in the same directions, the spinon lines which may be inserted  
are all chiral. The spinons are restricted to propagate only within a  
cone defined by two spatial quadrants (Fig.~\ref{lightcone}), and no closed  
spinon paths are possible in such states. The same is true for the opposing  
``ferroelectric'' state, the state full of lines, in which the string  
created between the two spinons appears as an antiline.  
 
As suggested by the simple example of the previous paragraph,  
the spinon motion can be determined from the line rules. The trace left  
by two spinons which are created on a common tetrahedron and move away  
from each other corresponds, in the line representation, to a string  
which changes the state of the underlying link: the string creates a line  
segment if the link was empty and annihilates the corresponding segment if  
it is already present. Following the rules of Fig.~\ref{rules}, if the line  
segment created by the string crosses a preexisting line, the resulting  
state on the corresponding tetrahedron is that indicated as 2 in  
Fig.~\ref{rules}. Once again we show for clarity in Figs.~\ref{spinon}(c)  
and (d) the line representations of the dimer/spinon states depicted in  
Figs.~\ref{spinon}(a) and (b) for the checkerboard lattice. In the general  
case of non--uniform states with a finite density of lines, the spinon  
strings are no longer constrained to be chiral (Fig.~\ref{spinon}). When  
the string of a propagating spinon collides with a six--vertex line  
representing the dimer configuration, the lines (and hence the spinon  
path) are effectively reflected. 
 
\section{Thermally Driven Deconfinement Away From the Klein Point} 
\label{thermalklein}
 
As noted above, the dipolar correlations characterizing the deconfined  
Klein critical point are driven by classical fluctuations, whence the  
most appropriate designation is as a classical $T = 0^+$ critical point. 
Away from $T = 0$, the critical thermal fluctuations of the dimer coverings 
produce an entropic spinon--spinon interaction. This interaction may be  
determined by replacing each spinon with a static monomer inserted  
in the dimer covering, reducing the problem to the classical statistical  
mechanics of close--packed dimer coverings \cite{Kasteleyn63, Fisher63}.  
This reduction makes use of the fact that the singlet coverings are  
linearly independent \cite{note4}. The critical dipolar fluctuations  
of the underlying dimer field produce an effective ``Coulomb''  
interaction between the spinons \cite{dipole}, which in the pyrochlore  
lattices under consideration arises from the local conservation law of  
zero divergence.  
 
We will demonstrate the existence of the Coulomb phase for the checkerboard  
lattice; a similar analysis may be applied for the 3d pyrochlore case. 
The derivation of the effective Coulomb interaction begins by assigning a  
charge to each spinon, for which it is convenient to expand the six--vertex  
representation in order to include states containing two spinons. It is  
important to note that spinons are created in pairs (singlet--triplet  
excitation), and each pair generates a single crossed plaquette of higher  
energy, because is possesses no singlets, which will be called a ``defect''.  
When the spinons propagate away from each other they are also separated  
from the defect (Fig.~\ref{spinon}), and preservation of the ice--rules  
requires that the number of defects does not increase. We will show below  
that the effective charge $q_d$ of the defect is opposite in sign to the  
effective charge $q_s$ of the spinons, with magnitude $q_d = - 2 q_s$.  
[Superficially, this overall neutrality of three quasiparticles forms a  
geometric analog of the quark content of the neutron, whose basic quark  
structure is $|u d d \rangle$ with the charges of the $u$ and $d$ quarks  
being respectively $(2e/3)$ and $(-e/3)$.]  
 
\begin{figure}[t!]
\vspace*{-0.5cm}
\hspace*{0.3cm}
\includegraphics[angle=90,width=10cm]{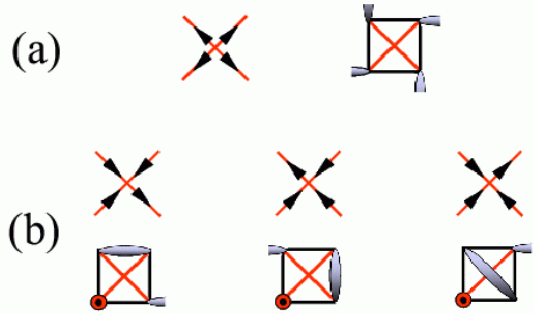}
\vspace{-2.7cm}
\caption{ (Color online.) (a) Vertex representation of a crossed
plaquette without any singlet (defect). (b) Vertex representation for 
spinons that are located in one sublattice of the checkerboard lattice.}
\label{spinvert}
\end{figure}

To represent states with two spinons, we will include the four additional 
vertex configurations depicted in Fig.~\ref{spinvert}. Note that these 
configurations have non--zero divergence, in contrast to the six--vertex 
states (Fig.~\ref{rules}) used to represent the dimer coverings (which  
are the vacuum of charge). According to the previous rules, the defect  
is represented by four outgoing arrows [Fig.~\ref{spinvert}(a)]. For 
the checkerboard geometry it is convenient to introduce the two N\'eel  
sublattices A and B of the underlying square lattice; a number of  
equivalent definitions may be found for a two--sublattice decomposition  
of the 3d pyrochlore. A one--to--one mapping between states containing  
spinons and the vertex configurations may be found by restricting the  
spinons to one of the sublattices (for example A). A
complete classification of all spinon configurations 
may be obtained by introducing an additional Ising--like quantum number
which specifies on which of the two sublattices our plaquettes are located.
Such an additional tabulation is not necessary for 
the examination of the long distance interaction.
We may rely instead only on the consequences  
of the vertex representation illustrated in Fig.~\ref{spinvert} for
plaquettes of one sublattice, which are  
sufficient for determining the asymptotic (long--distance) form of the  
interspinon interaction. The fact that spinons may propagate from one 
sublattice to the other due to the existence of diagonal bonds in dimer 
coverings of the checkerboard lattice is nevertheless crucial. This is 
not the case for quantum dimer models on bipartite lattices, whose dimers 
always link the two different sublattices. 
We remind the reader that on a bipartite lattice each spinon may be  
assigned a unique effective charge given by the sublattice parity of  
the spinon location (usually +1 on one sublattice, and -1 on the other).  
This simple assignment is invalid on non--bipartite lattices such as the  
pyrochlore, where the additional range of possible spinon configurations  
necessitates an essential extension of the definition of the effective  
spinon charge. 

Because the spinons occupy sites of the original lattice, their position 
will be denoted by one arrow on the dual lattice. Specifically, we take 
this as the ingoing arrow coming from the lower left corner of a plaquette 
with only one outgoing arrow [Fig.~\ref{spinvert}(b)]. If one considers  
vertex states such that for each vertex with maximal divergence  
[Fig.~\ref{spinvert}(a)] there are two with only one outgoing arrow  
[Fig.~\ref{spinvert}(b)], the mapping between the two representations 
is one--to--one (under the condition that all the spinons occupy the  
same sublattice). The Coulomb interaction and the effective charge of  
the spinons emerge from the standard procedure used to map the ice  
problem onto a compact U(1) gauge theory \cite{Kogut}. Here we follow  
the derivation and notation of Ref.~\cite{CF} to describe this procedure  
and derive the effective charges of defects and spinons.  
 
The local zero--divergence condition defining the dimer coverings on the  
checkerboard and pyrochlore lattices implies that these states can be  
considered as the physical Hilbert space of a compact $U(1)$ gauge theory.  
The local $U(1)$ gauge invariance becomes more explicit if one defines  
pseudospin--1/2 operators ${\cal S}^z_i$ on each site $i$ of the lattice.  
The dual lattice formed by the centers of the tetrahedrons (vertices) can  
be divided into two sublattices ${\cal A}$ and ${\cal B}$. The site  
coordinate $i$ of the original lattice denotes the links of the dual  
lattice. ${\cal S}^z_i = 1/2$ if the arrow on link $i$ points from the  
${\cal A}$ to the ${\cal B}$ sublattice, ${\cal S}^z_i = - 1/2$ if the  
arrow points from ${\cal B}$ to ${\cal A}$. The zero--divergence condition  
can be expressed in terms of these variables as 
\begin{equation} 
\sum_{i \in \boxtimes} {\cal S}^z_i = 0, 
\end{equation} 
and the operators  
\begin{equation} 
{\cal U}_{\alpha} = e^{i \phi_{\alpha} \sum_{i \in \alpha} {\cal S}^z_i} 
\end{equation} 
are the local U(1) symmetry transformations associated with this 
conservation law. We note that the conservation law is a property of 
the Hilbert space and is therefore present for any theory defined on 
this space. An equivalent constrained low--energy Hilbert space was 
derived for an XXZ model on a pyrochlore lattice in 
Ref.~\cite{ring_exchange1}.
 
We present in some detail the consequences of the zero--divergence 
condition for the checkerboard lattice ($d = 2$), for which the 
notation is slightly more compact; the results we obtain are unchanged  
for the pyrochlore lattice ($d = 3$). The $U(1)$ gauge structure  
of the Hilbert space indicates that one may gain further insight by  
identifying the variables playing the role of the electric field and  
of the vector potential in electrodynamics. For this purpose, it is  
convenient to introduce quantum rotor variables and the coordinate  
${\bf r}= n {\bf e}_1 + m {\bf e}_{2}$ to span the dual lattice (the  
lattice of centers of the tetrahedrons). The primitive vectors ${\bf e}_1$  
and ${\bf e}_{2}$ define the two perpendicular directions on the dual  
lattice. On each link of this lattice we define an angular--momentum  
variable ${\tilde l}_{\mu}({\bf r}) = \pm 1/2$, where $\mu({\bf r})$  
indicates the oriented link from ${\bf r}$ to a neighboring dual--lattice  
site ${\bf r} + {\bf e}_{\mu}$. The state ${\tilde l}_{\mu}({\bf r}) = 1/2$ 
(${\tilde l}_{\mu}({\bf r}) = -1/2$) corresponds to an arrow parallel  
(anti--parallel) to the relative vector ${\bf e}_{\mu}$. With this  
definition, ${\tilde l}_{\mu}(\bf r)$ changes sign under inversion:  
${\tilde l}_{-\mu}({\bf r}) = - {\tilde l}_{\mu}({\bf r})$ ($-\mu({\bf  
r})$ indicates the oriented link from ${\bf r}$ to ${\bf r}-{\bf e}_{\mu}$).  
For convenience we introduce the variable $l_{\mu}({\bf r}) = {\tilde  
l}_{\mu}({\bf r}) + 1/2$, which takes the integer values 0 or 1. We  
employ the discrete gradient 
\begin{equation} 
\nabla_{\mu}  l_\nu({\bf r})= l_\nu({\bf r})-l_\nu({\bf r}-{\bf e}_\mu) 
\end{equation} 
to reexpress the ice rules as an explicit zero--divergence condition on 
the vector field $l_{\mu}({\bf r})$, 
\begin{equation} 
\sum_{\mu = 1,2} \nabla_{\mu} l_\mu({\bf r}) = 0. 
\label{Gauss} 
\end{equation} 
Equation (\ref{Gauss}) corresponds to Gauss' law for electrodynamics in  
the absence of external charges, $\nabla \cdot {\bf E} = 0$. Vertex 
configurations violating the ice rules, such as those depicted in 
Fig.~\ref{spinvert}, should then carry an ``effective charge'' $Q =  
\sum_{\mu = 1,2} \nabla_{\mu} l_\mu({\bf r})$. Here $Q = \sum_{\mu =  
1,2} \nabla_{\mu} l_\mu({\bf r}) = 2$ for the configurations of  
Fig.~\ref{spinvert}(a) (defects) and $Q = \sum_{\mu = 1,2} \nabla_{\mu}  
l_\mu({\bf r}) = -1$ for the three 
configurations of Fig.~\ref{spinvert}(b)  
(spinons): the spinons carry a positive unit charge while the defects 
carry two units of negative charge. The ``electric field'' is then 
defined as the operator ${\bf E}_{\mu} ({\bf r})$  with eigenstates  
$| l_\mu({\bf r}) \rangle$ and eigenvalues $l_\mu ({\bf r})$. The 
canonically conjugate operator $\Theta_{\nu}({\bf r})$ satisfying  
\begin{equation} 
[\Theta_{\nu}({\bf r'}), {\bf E}_{\mu} ({\bf r})]=  
i \delta_{{\bf r},{\bf r'}} \delta_{\mu,\nu}  
\end{equation} 
plays the role of the vector potential. Because in this section we focus  
only on the high--temperature ($T \rightarrow \infty$) fixed point whose 
properties are determined solely by the gauge structure of the Hilbert 
space, we will not discuss any particular quantum gauge theory. The above  
derivation is used only to identify which variables play the role of the  
electric field and the charge. 
 
The Klein point at $T = 0^+$ corresponds to the infinite--temperature fixed 
point. Such a point is characterized by a deconfined two--component (spinons 
and defects) Coulomb gas. The spinons exist only as gapped excitations  
(of gap $\Delta \propto J$), and the spinon--spinon interaction is then  
logarithmic, $V(r)= \gamma \sum_{j<k} q_j q_k \ln{(r_{jk}/a)}$, for $d =  
2$ and a power law, $V(r) = {\tilde \gamma} \sum_{j<k} q_j q_k /{r_{jk}}$,  
for $d = 3$. The spinon--defect interaction has the opposite sign  
(attractive) and its absolute value is two times greater. Although a  
spinon--defect pair is critically confined in the $d = 2$ case by the  
logarithmic interaction, the prefactor $\gamma$ is generally sufficiently  
small that there is a divergence in the expectation value of $\langle r^2  
\rangle$, which implies a divergent ``dielectric constant'' \cite{Krauth03}.  
At finite temperatures there is a finite concentration $\rho \sim e^{- \beta  
\Delta}$ [$\beta = (k_{\rm B} T)^{-1}$] of thermally excited spinons in  
equilibrium, and the effective spinon--spinon or spinon--defect 
interaction is screened, decaying exponentially for distances $r  
\gg 1/\rho^{1/d}$. However, the critical behavior remains observable  
for $r \ll 1/\rho^{1/d}$.  
 
For general $K \neq K_{c}$ and $|K - K_c| \ll J $, the energy difference  
between the singlet and triplet states on each tetrahedron is much  
smaller than the gap to the quintet state ($S = 2$). \cite{012_note} 
It is clear that for $\epsilon = |K - K_{c}| \ll J$ the degeneracy 
between the singlet and triplet states is effectively restored at 
temperatures $T$ such that 
\begin{equation} 
\epsilon \ll k_{\rm B} T \ll J.   
\label{glassbound} 
\end{equation} 
In this case, only the manifold of singlet dimer  
coverings found as the ground states at $K = K_{c}$ is accessible. Thus the 
effect of a finite temperature on a system in the proximity of a Klein point  
is the effective restoration of extensive configurational entropy, critical  
dipolar correlations, topological order, and spinon deconfinement. 
 
We comment that the appearance of deconfinement at high temperatures,  
also in systems far more general than those considered here, is hardly  
surprising. We repeat well known facts: various ordered phases generally  
melt via the appearance of topological defects; these defects become  
deconfined at the melting temperature; the origin of deconfinement is  
an energy--entropy balance; entropic effects modify the effective  
interactions between the defects, leading to a vanishing interaction  
at the onset of deconfinement. A careful treatment of entropic effects  
often involves detailed contour--counting arguments \cite{Peierls}.  
The effective force between topological defects in a wide range of  
systems, including vortices in superconductors and dislocations 
and disclinations in elastic media, is universally expected to be  
algebraic, as may be seen directly from duality arguments \cite{Kleinert,  
XY_duality,elastic_duality}. In the current context the topological  
defects correspond to monomers or spinons in the dimer coverings and the  
Coulomb field lines to the many possible trajectories for the motion of a  
defect. Pure Coulomb interactions vanish at large separations in dimensions  
$d > 2$, suggesting the possibility of high--dimensional fractionalization.  
The peculiar aspect of the pyrochlore systems under consideration is that  
the critical point separating the low--temperature confined state from the  
high--temperature deconfined phase may be driven to $T = 0^{+}$ by optimizing  
the frustration in $H$. This is the singular characteristic of the Klein  
point: the exponentially large ground--state sector is defined solely by 
the local zero--divergence constraint. The system is then driven only by 
entropic fluctuations, which in combination with the local constraint 
give rise to the classical critical behavior.  
 
{\section{Zero--temperature states away from the Klein point}} 
 
\subsection{Local perturbations in the non--orthogonal dimer basis} 
 
In this section we examine explicitly the possible ground states away from  
the Klein point. To this end, we begin by developing a systematic procedure  
for evaluating ``matrix elements'', $\langle \psi_a |{\hat {\cal O}}| \psi_b  
\rangle$, within the non--orthogonal basis of  
singlet dimer coverings which constitute the low--energy sector in the  
vicinity of the Klein point. We note that because of the non--orthogonality  
a standard (bra--operator--ket) expectation value computed between two  
states in the manifold is not a true matrix element, and is referred to  
henceforth as a ``bracket''. We will illustrate the calculational procedure  
and results for the checkerboard lattice, and state only that their  
generalization to the pyrochlore is straightforward (if lengthy). 
 
The evaluation procedure consists of elementary rules which are applied  
to the graph associated with the bracket to be evaluated, and it is a  
generalization of the standard method used to compute the overlap between  
two singlet dimer coverings on bipartite lattices \cite{Eduardo, earlierRVB, 
RK, kevin}. We first extend this method to compute the overlap between  
non--bipartite singlet coverings, because on the pyrochlore system the  
dimers are not necessarily formed from sites on different sublattices.  
Secondly and more importantly, we provide straightforward rules for  
computing brackets of spin--product operators of Heisenberg type,  
${\vec S}_i \cdot {\vec S}_j$. Brackets of products of Heisenberg spin  
operators, as in the second term of Eq.~(\ref{esh}), can be evaluated  
by a simple extension of these rules. 
 
In this section, we will consider a perturbation which is proportional to  
the first term of Eq.~(\ref{esh}), a Heisenberg interaction between any  
pair of sites in the same crossed plaquette 
\begin{equation} 
H_{1} = \Delta J \sum_{\langle ij \rangle, \alpha} {\vec S}_i^{\alpha}  
\cdot {\vec S}_j^{\alpha}, 
\label{Heisenberg_perturbation} 
\end{equation} 
such that after the perturbation $K \neq 4J/5$. For the Klein model on the  
square lattice it is transparent that all states within the ground--state  
basis remain degenerate to first order in the perturbation $H_1$.  
This statement follows from two observations: the diagonal brackets,  
$\langle \psi_a | H_1 | \psi_a \rangle$, are the same for any dimer  
state $|\psi_a \rangle$; the off--diagonal terms, $\langle \psi_a  
|{\hat {\cal O}}| \psi_b \rangle$, with $a \neq b$, are zero for  
any local operator ${\hat {\cal O}}$, in particular for ${\hat  
{\cal O}} = H_1$. Thus the degeneracy is lifted only to second  
order in $H_1$ on the square lattice. The situation is completely  
different for the same model on the pyrochlore and checkerboard  
lattices: because in these cases the ground--state degeneracy is  
extensive at the Klein point, two different ground states can be connected  
by local operators. As a consequence the off--diagonal brackets do not in  
general vanish, and the degeneracy is lifted to first order in the  
perturbation.  
 
The usual approach to this problem is to neglect the non--orthogonality  
of the singlet dimer coverings and propose a minimal quantum dimer model  
which includes only processes of the Rokhsar--Kivelson (RK) type \cite{RK}.  
We note, however, that this procedure does not lead to the RK model in the  
pyrochlore geometries under consideration because, although the Hamiltonian  
is the same (a U(1) gauge magnet \cite{Horn}), the zero--divergence or gauge  
condition which defines the ``physical'' Hilbert spaces is different. In what  
follows we depart from this conventional treatment, and show that it is  
indeed incorrect for the case of interest. The non--orthogonality of the  
dimer basis introduces subtleties not directly evident from the conventional  
approach, as a consequence of which we will show that the valence--bond  
crystal (VBC) orderings chosen by tunneling events differ from the ones  
expected for a theory which includes only processes of the RK type.  
 
\begin{figure}[t!]
\vspace*{-0.7cm}
\hspace*{-0.7cm}
\includegraphics[angle=90,width=9cm]{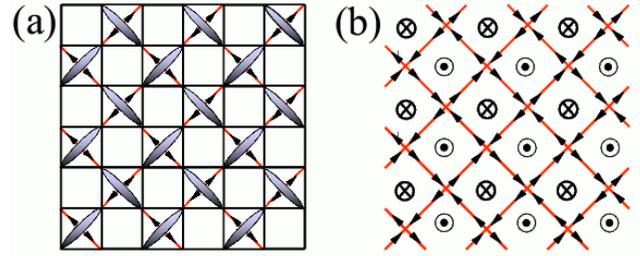}
\vspace{-5.0cm}
\caption{ (Color online.) The ``antiferroelectric'' state. Dots and 
crosses at the centers of the plaquettes of the dual lattice indicate 
the ``flux'' associated with the arrow circulation on the corresponding 
plaquette. The two antiferroelectric states are the only ones for which 
each plaquette of the dual lattice has a net chirality.}
\label{afes}
\end{figure}

In terms of the arrows (the six--vertex representation) shown in  
Fig.~\ref{afes}, the antiferroelectric  
state becomes a staggered flux phase in which each square plaquette of  
the dual lattice has a well--defined chirality (all of its arrows circulate  
in the same way) and the direction of circulation alternates from clockwise  
to counterclockwise. With this starting state one may invert the sign of  
the current (direction of the arrows) around any plaquette. This is the  
elementary process, or corner flip in the line representation, depicted  
in Fig.~\ref{fig:pr1}(b). The antiferroelectric state allows for the  
largest number of these elementary processes, and each can be interpreted  
as the creation of a localized defect on an uncrossed plaquette of the  
antiferroelectric background. The tunneling between states with defects  
in different positions lowers the energy of the system. While the creation  
of one of these defects is indeed the elementary RK process normally  
invoked as the most relevant one in these cases \cite{Raman05}, we will  
show that its amplitude is exactly equal to zero for the perturbation  
$H_1$. 
 
To first order in the perturbation the state of lowest energy is the  
linear combination of singlet dimer coverings which minimizes $H_1$.  
Up to an irrelevant constant, the Hamiltonian $H_1$ can be rewritten  
in the more convenient form  
\begin{equation} 
H_{1} = {\textstyle \frac{1}{2}} \Delta J \sum_{\langle i j \rangle,  
{\alpha}} P^{\alpha}_{ij} , 
\end{equation} 
where $P^{\alpha}_{ij}$ is the permutation operator which interchanges  
the spins on sites $i$ and $j$ of plaquette $\alpha$. The diagonal  
brackets are manifestly the same for all singlet dimer coverings.  
Specifically, on the checkerboard lattice and for any system with PBCs, 
\begin{equation} 
E_d = \langle \psi_{a}| H_{1}| \psi_{a} \rangle = {\textstyle \frac{3}{8}}  
N \Delta J, 
\label{uniform} 
\end{equation} 
with $N$ the number of lattice sites and $|\psi_{a} \rangle$ denoting any  
dimer covering satisfying the Klein constraint (all crossed plaquettes  
contain one dimer). As noted in Sec.~III, systems with OBCs possess   
additional ground states containing crossed plaquettes with two dimers or 
(one or two) spinons, and  
these would have a different diagonal energy but are not thermodynamically  
relevant. We focus henceforth on ground states satisfying PBCs. The  
degeneracy of dimer coverings is lifted only by superposing different  
allowed coverings with finite amplitudes, {\it i.e.}~by considering states  
of the form 
\begin{equation} 
|\psi \rangle = \sum_{a} c_{a} |\psi_{a} \rangle, 
\label{superposition} 
\end{equation} 
with the normalization condition 
\begin{equation} 
\sum_{k} |c_{k}|^{2} + \sum_{l \neq k} c^{*}_{l} c_{k} \langle  
\psi_l | \psi_k \rangle = 1. 
\label{norm} 
\end{equation} 
 
\begin{figure}[t!]
\vspace*{-0.3cm}
\hspace*{-0.8cm}
\includegraphics[angle=90,width=9cm]{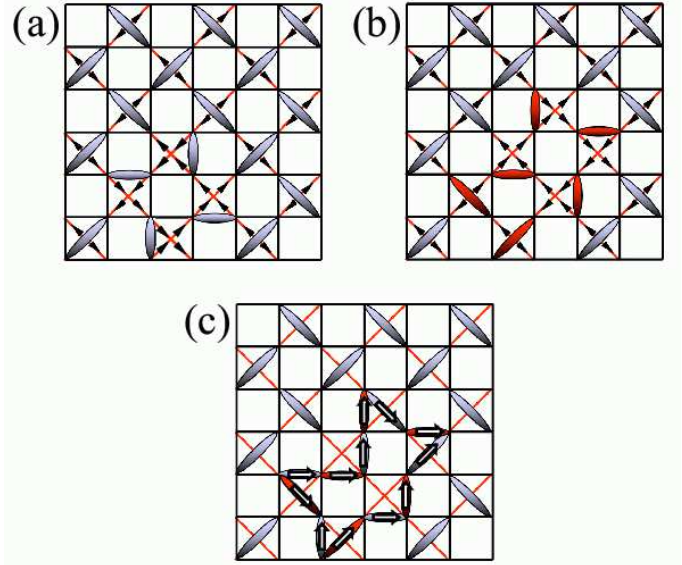}
\vspace{-0.8cm}
\caption{ (Color online.) Example (c) of the graph $G^{ab}$ which results 
from superposing two different dimer coverings [(a) and (b)]. These dimer 
coverings correspond to an antiferroelectric background with a single 
elementary process operating on different (nearest--neighbor) uncrossed 
plaquettes.}
\label{graph}
\end{figure}
 
\subsection{Loop Structure and Rules} 
 
We proceed by listing a complete set of rules for the evaluation 
of brackets in the dimer basis. These general, purely  
topological rules extend early and pioneering ideas concerning  
resonating valence--bond states, and initial results obtained by using  
these concepts for square lattices \cite{Eduardo,earlierRVB, RK, kevin}. 
The normalization of $|\psi \rangle$ and computation of $\langle \psi |  
H_1 |\psi \rangle$ require evaluating the overlaps $\langle \psi_{a}|  
\psi_{b} \rangle$ and brackets of the form $\langle \psi_{a}| P_{ij}|  
\psi_{b} \rangle$, with $|\psi_{a} \rangle$ and $| \psi_{b} \rangle$  
any states in the dimer basis. To compute the overlap we first assign a  
given orientation to each singlet dimer, indicated with arrows pointing  
upwards for vertical dimers and to the right for diagonal and horizontal  
bonds (Fig.~\ref{graph}). [We remark that ``vertical'' and ``horizontal'' 
refer here to the real lattice, and not to the rotated lattice used in  
the discussion of the six--vertex and line representations in Sec.~III.]  
With this convention, a superposition of dimer configurations corresponding  
to $|\psi_{a} \rangle$ and $|\psi_{b} \rangle$, with the differing bonds  
marked as in Fig.~\ref{graph}(c) (bonds which do not differ give trivial  
loops of length two, introducing factors of unity), forms a graph $G^{ab}$.  
In general, this graph may consist of more than one disconnected loop  
$\Gamma^{ab}_j$.  
 
Let $L(\Gamma^{ab}_j)$ be the length of the loop $\Gamma^{ab}_j$ (the  
number of singlet dimers forming the loop), $N_{\Gamma}(G^{ab})$ be  
the number of disconnected loops in graph $G^{ab}$, and $n_c$ be the  
number of arrows in the graph which circulate clockwise in their loop.  
With these definitions the overlap 
$\langle \psi_{a}| \psi_{b} \rangle$  
may be expressed as 
\begin{equation} 
g_{ab}=\langle \psi_{a}| \psi_{b} \rangle = (-1)^{n_c + L(G^{ab})/2}  
2^{N_{\Gamma}(G^{ab}) - L(G^{ab})/2}, 
\label{gab} 
\end{equation} 
where  
\begin{equation} 
L(G^{ab}) = \sum_{j=1,N_{\Gamma}(G^{ab})} L(\Gamma^{ab}_j) 
\end{equation} 
is the total length of the graph. The determination of the non--trivial  
brackets $\langle \psi_{a}| P_{ij}| \psi_{b} \rangle$ proceeds by examining  
how the length and dimer orientation of the graph $G^{ab}$ are affected  
when the permutation operator $P_{ij}$ is applied to the state $| \psi_{b}  
\rangle$. The resulting graph will be denoted as $G^{ab}_{ij}$. The  
application of $P_{ij}$ may alter the loop geometry in only a restricted  
set of simple ways, of which we next provide a systematic list. 
 
\renewcommand\theenumi{\roman{enumi}}  
\renewcommand\labelenumi{[{\bfseries\theenumi}]} 
 
\begin{figure}[t!]
\vspace*{-0.3cm}
\hspace*{-0.8cm}
\includegraphics[angle=90,width=9cm]{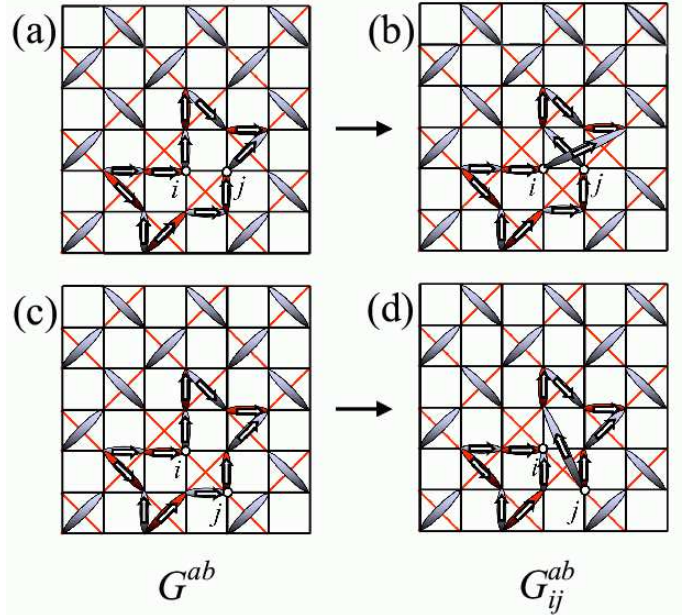}
\vspace{-0.8cm}
\caption{ (Color online.) Effect of the permutation operator $P_{ij}$ on 
graph $G^{ab}$ (Fig.~\ref{graph}) when sites $i$ and $j$ belong to the same 
loop of $G^{ab}$ and their separation is odd [case (a) $\rightarrow$ (b)] or
even  [case (c) $\rightarrow$ (d)].}
\label{PijtypeI}
\end{figure}

\begin{enumerate} 
 
\item \label{l1} If $i,j \in G^{ab}$ and they belong to the same loop 
$\Gamma$, there are two possibilities:  
 
\begin{enumerate} 
 
\item \label{l11} 
when the distance (number of dimers) on the loop between sites $i$ and  
$j$ is an odd number, the effect of $P_{ij}$ is to cut the loop into two  
segments and to reconnect them while exchanging the ends of one of the  
segments [Figs.~\ref{PijtypeI}(a) and (b)]. The length of  
the loop under consideration is unaltered in this process. The only  
effective modification is the reversal of an odd number of arrows,  
causing a sign--change relative to $\langle \psi_{a}| \psi_{b} \rangle$,  
\begin{equation} 
\langle \psi_{a}| P_{ij}| \psi_{b} \rangle = - g_{ab}. 
\end{equation} 
 
\item \label{l12} 
 
when the distance between sites $i$ and $j$ is even, $P_{ij}$ divides the  
loop $\Gamma$ into two smaller loops, $\Gamma_1$ and $\Gamma_2$, preserving  
the total length, $L(\Gamma) = L(\Gamma_1) + L(\Gamma_2)$  
[Figs.~\ref{PijtypeI}(c) and (d)]. The circulation of the  
arrows is unaltered, and thus from Eq.(\ref{gab}) the net change relative  
to graph $G^{ab}$ is a supplementary factor of 2 due to the increase in  
the number of loops, $N_{\Gamma}(G^{ab}_{ij}) = N_{\Gamma}(G^{ab}) + 1$,  
whence 
\begin{equation} 
\langle \psi_{a}| P_{ij}| \psi_{b} \rangle = 2 g_{ab}. 
\end{equation} 
 
\end{enumerate}   
 
\item \label{l2}  
If only one of the two sites ($i$ or $j$) belongs to graph $G^{ab}$, 
the effect of $P_{ij}$ is to increase the length of the graph by two  
links, $L(G^{ab}_{ij}) = L(G^{ab}) + 2$. There is no change of sign 
because the arrows of the two additional bonds circulate in opposite  
directions relative to their point of contact. From Eq.~(\ref{gab}) the 
net result relative to graph $G^{ab}$ is a factor of $1/2$ due to 
the length increase, and so 
\begin{equation} 
\langle \psi_{a}| P_{ij}| \psi_{b} \rangle = {\textstyle \frac{1}{2}}  
g_{ab}. 
\end{equation} 
 
\item \label{l3}  
If neither $i$ nor $j$ belong to graph $G^{ab}$, there are two 
possible cases:  
 
\begin{enumerate} 
 
\item \label{l31}  
if there is a singlet bond connecting the sites $i$ and $j$,  
\begin{equation} 
\langle \psi_{a}| P_{ij}| \psi_{b} \rangle =  - g_{ab}; 
\end{equation} 
 
\item \label{l32} 
if the sites $i$ and $j$ are not connected by a singlet bond,  
\begin{equation} 
\langle \psi_{a}| P_{ij}| \psi_{b} \rangle = {\textstyle \frac{1}{2}}  
g_{ab}. 
\end{equation} 
 
\end{enumerate} 
 
\item \label{l4}   
Finally, if $i,j \in G^{ab}$ but the sites belong to different loops 
$\Gamma_i$ and $\Gamma_j$, the effect of $P_{ij}$ is to fuse both loops 
into a single loop $\Gamma_{ij}$. Because the total length, $L(\Gamma_{ij}) 
 = L(\Gamma_i) + L(\Gamma_j)$, and the sign are preserved under this  
operation, the net result is a factor $1/2$ which arises from the decrease  
by one in the number of loops, $N_{\Gamma}(G^{ab}_{ij}) = N_{\Gamma}(G^{ab}) 
 - 1$, leading to   
\begin{equation} 
\langle \psi_{a}| P_{ij}| \psi_{b} \rangle = {\textstyle \frac{1}{2}}  
g_{ab}. 
\end{equation} 
 
\end{enumerate} 
 
These rules allow a very straightforward derivation of Eq.~(\ref{uniform}): 
because $G^{aa}$ is the null graph, the effect of $P_{ij}$ corresponds 
to case [\ref{l3}]. For any dimer covering there are $N/2$ bonds of 
type [\ref{l31}] and $5N/2$ bonds of type [\ref{l32}], and adding the  
two contributions yields 
\begin{equation} 
E_d = \langle \psi_{a}| H_{1}| \psi_{a} \rangle = \frac{1}{2} \Delta J  
\left( - g_{aa} \frac{N}{2} + \frac{g_{aa}}{2} \frac{5N}{2} \right). 
\end{equation} 
From $g_{aa} = 1$ [see Eq.~(\ref{gab})] we recover Eq.~(\ref{uniform}).  
 
With the same rules one may compute the energy of the most general  
state within the Klein--point ground--state manifold whose expression is 
given by Eq.~(\ref{superposition}), 
\begin{equation} 
E  = E_d + \frac{1}{2} \Delta J \sum_{ l \neq k } c^{*}_{l} c_{k} 
\left[ g_{lk} \left( \frac{3 L(G^{lk})}{4} \! - \! \frac{|{\cal{V}}^{lk}|}{2}  
\right) + W_{lk} \right]. 
\label{general_energy} 
\end{equation} 
Here $W_{lk} \equiv 2 n^{lk}_{ib}- n^{lk}_{ia}$, where $n^{lk}_{ia}$ 
($n^{lk}_{ib}$) is the number of pairs $\langle i,j \rangle$  
which are of type [\ref{l11}] ([\ref{l12}]) relative to the  
graph $G^{lk}$, while ${\cal{V}}^{lk} = n^{lk}_{ia} + n^{lk}_{ib}$  
is the total number of pairs $\langle i,j \rangle$ of type [i].  
The reason why 
only cases of type [i] need be considered in Eq.~(\ref{general_energy}) 
is that terms of types [\ref{l2}], [\ref{l32}], and [\ref{l4}] all give 
the same contribution, $g_{ab}/2$.  
 
\begin{figure}[t!]
\vspace*{-0.5cm}
\hspace*{-0.8cm}
\includegraphics[angle=90,width=9cm]{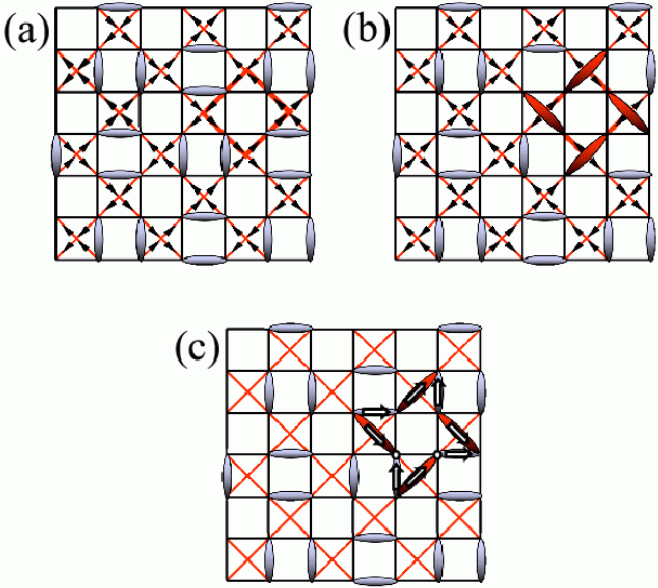}
\vspace{-1.0cm}
\caption{ (Color online.) Graph $G^{od}$ resulting from the superposition 
of two dimer coverings which differ minimally. The physical process 
connecting these is the RK--type or corner--flip process represented 
in Fig.~\ref{fig:pr1}.}
\label{RK}
\end{figure}
  
\subsection{Energy Calculations} 
 
Next we apply the rules of Subsec.~VB to the checkerboard lattice. 
Equation (\ref{general_energy}) allows one to compute the energy of any  
variational state. Its second term corresponds to the energy change  
$\delta E$ due to the linear combination of different singlet dimer  
coverings. The magnitude of this change therefore reflects the degree  
of ``mixing'' between different dimer coverings induced by $H_1$. To  
develop an intuitive understanding we begin by considering the linear  
combination of two dimer coverings, a natural choice being to superpose  
two coverings whose dimer configurations differ minimally (such as the  
two states depicted in Fig.~\ref{fig:pr1}), {\it i.e.}~by the RK or  
corner--flip process discussed above. We remind the reader that the  
six--vertex representation of this process entails a reversal of the  
arrows on a single plaquette of the dual lattice with a well--defined  
chirality [Figs.~\ref{RK}(a) and (b)]. If $| \psi_o \rangle$ and  
$| \psi_d \rangle$ are any such pair of dimer coverings which differ  
minimally, their overlap is  
\begin{equation} 
g_{od} = \langle \psi_o | \psi_d \rangle = 2 \times 2^{-4} = {\textstyle  
\frac{1}{8}}, 
\label{overl} 
\end{equation} 
because $L(G^{od}) = 8$ and $n_c = 4$ [Fig.~\ref{RK}(c)]. To obtain the  
energy of the normalized state 
\begin{equation} 
| \psi_{od} \rangle = \alpha| \psi_{o} \rangle + \beta |\psi_{d}\rangle, 
\label{N-d} 
\end{equation} 
one replaces the coefficients of Eq.~(\ref{general_energy}) with the  
values dictated by the above rules. The number of pairs $\langle i,j 
\rangle$ with both sites on the graph is $|{\cal{V}}^{od}| = 12$,  
composed of 8 pairs of type [\ref{l11}] and 4 of type [\ref{l12}],  
as a result of which $W^{od} = 0$. Replacing these quantities in 
Eq.~(\ref{general_energy}) gives  
\begin{equation} 
\langle \psi_{od}| H_1 | \psi_{od} \rangle = E_d. 
\label{ERK} 
\end{equation} 
 
The fact that the energy is unchanged ($\delta E = 0$) by taking the  
linear combination (\ref{N-d}) indicates that the elementary RK--type  
process is not generated by $H_1$. We note that all pairs of sites $\langle  
i,j \rangle$ in the only loop of the graph $G^{od}$ [Fig.~\ref{RK}(c)] are  
separated by a distance on the loop no greater than 2; loops satisfying  
this condition will be defined as {\sl simple}. As an example, the loop  
depicted in Fig.~\ref{graph}(c) is not simple because it contains site  
pairs $\langle i,j \rangle$ separated by five and six bonds.  
In general, $\delta E$ vanishes for any pair  
of dimer configurations, $| \psi_a \rangle$ and $| \psi_b \rangle$,  
such that all loops in the graph $G^{ab}$ are simple. This property  
follows immediately from Eq.~(\ref{general_energy}): if all loops are  
simple one has that $n^{ab}_{ia} = 2n^{ab}_{ib} = L(G^{ab})$, 
implying $W_{ab} = 0$ and $|{\cal{V}}^{ab}| = 3L(G^{ab})/2$.  

\begin{figure}[t!]
\vspace*{-0.8cm}
\hspace*{-0.8cm}
\includegraphics[angle=90,width=9cm]{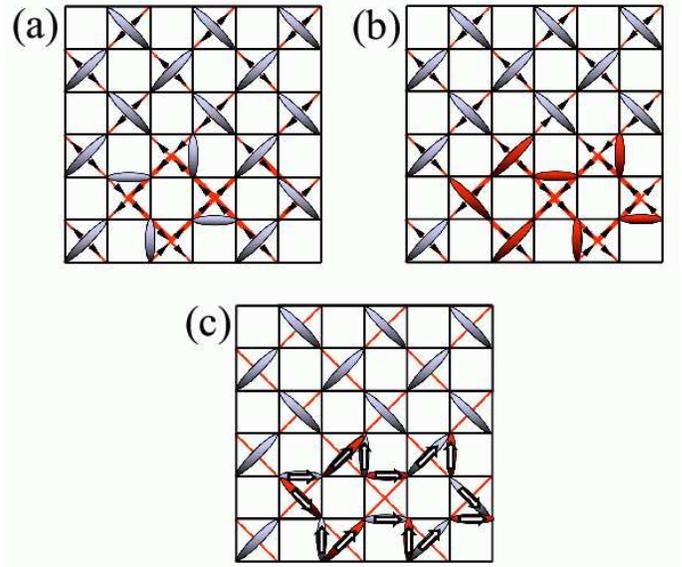}
\vspace{-1.0cm}
\caption{ (Color online.) Graph $G^{ab}$ resulting from the superposition of 
two states connected by inverting the arrows of two next--neighbor plaquettes 
of the dual lattice which have opposing, well--defined chiralities. The two 
plaquettes are highlighted in panels (a) and (b).}
\label{graph2}
\end{figure}
 
Because $H_1$ does not connect two minimally different dimer configurations  
[$L(G^{ab}) = 8$], it is necessary to seek the minimal relevant process,  
meaning that with the minimum value of $L(G^{ab})$, which is generated by  
$H_1$. The next possible value is $L(G^{ab}) = 12$ (no graphs exist with  
$L(G^{ab}) = 10$), an example of such a physical process being that  
connecting the states (a) and (b) in Fig.~\ref{graph}. In the six--vertex  
representation this process is permitted when all the arrows of a rectangle  
formed by two adjacent plaquettes of the dual lattice have the same  
chirality; this rectangle is highlighted in Figs.~\ref{graph}(a) and (b).  
The process consists of inverting the orientation of the six arrows, or  
the flux of the rectangle. For any pair of dimer coverings $| \psi_0  
\rangle$ and $| \psi_1 \rangle$ which differ by this process,  
\begin{equation} 
g_{01} = \langle \psi_0 | \psi_1 \rangle = - 2 \times 2^{-6} = 
- {\textstyle \frac{1}{32}},  
\end{equation} 
and for the normalized linear combination $| \psi_{01} \rangle = c_0|  
\psi_{0} \rangle + c_1 |\psi_{1}\rangle$ the energy change is  
\begin{equation} 
\langle \psi_{01}| H_1 | \psi_{01} \rangle = E_d + {\textstyle \frac{3}{64}}  
c_0 c_1 \Delta J . 
\label{E01} 
\end{equation} 
The minimum energy is obtained for $|c_0| = |c_1|$, which together with  
the normalization condition gives $c_0 = c_1 = 1/\sqrt{2(1 + g_{01})}$ for 
$\Delta J < 0$ and $c_0 = - c_1 = 1/\sqrt{2(1 - g_{01})}$ for $\Delta J >  
0$. The overlap is $|g_{01}| = \frac{1}{32} \ll 1$, giving $|c_0| = |c_1|  
\sim 1/\sqrt{2}$, as expected for the mixing of two orthogonal states  
with the same diagonal energy. This remains true for any pair of states  
connected by $H_1$ because $|g_{01}|$ is an upper bound for the absolute  
value of their overlap.  
 
In the next process to consider, the arrow directions on the perimeters  
of two next--neighbor plaquettes of the dual lattice are inverted if 
both plaquettes have opposing, well--defined chiralities, as indicated in  
the examples of Figs.~\ref{graph2}(a) and (b). In this case we denote the  
corresponding dimer coverings by $| \psi_0 \rangle$ and $| \psi_2 \rangle$,  
and the length of the associated graph is $L(G^{02}) = 14$. The overlap is  
$g_{02} = 1/64$ and the energy shift due to the perturbation for the 
linear combination $| \psi_{02} \rangle = c_0| \psi_{0} 
\rangle + c_2 |\psi_{2}\rangle$ is 
\begin{equation} 
\langle \psi_{02}| H_1 | \psi_{02} \rangle = E_d - {\textstyle  
\frac{3}{128}} c_0 c_2 {\Delta J}, 
\label{E01} 
\end{equation} 
with $c_0 = \pm c_2 = 1/\sqrt{2(1 \pm g_{02})}$ for $\Delta J = \pm  
|\Delta J|$.  
 
\begin{figure}[t!]
\vspace*{-0.8cm}
\hspace*{-0.8cm}
\includegraphics[angle=90,width=9cm]{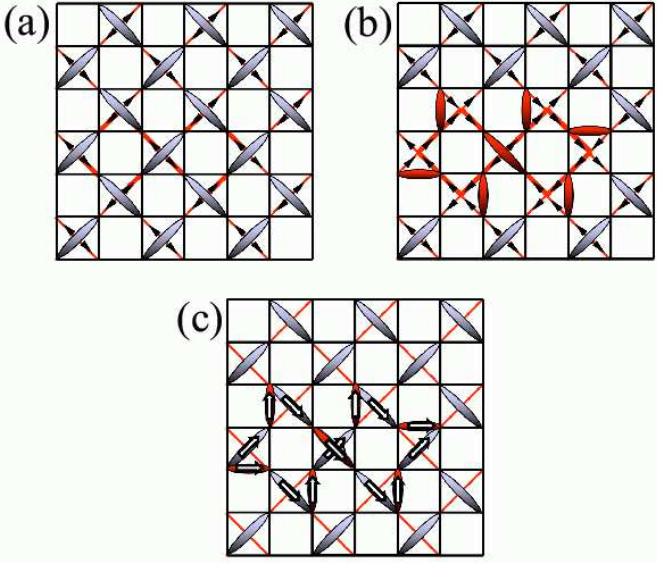}
\vspace{-1.0cm}
\caption{ (Color online.) Graph $G^{ab}$ resulting from the superposition of 
two states connected by inverting the arrows of two next--neighbor plaquettes 
of the dual lattice which have the same well--defined chirality. The two 
plaquettes are highlighted in panels (a) and (b).}
\label{twoloops}
\end{figure}

The other graph $G^{ab}$ of length $L(G^{ab}) = 14$ corresponds 
to a process which is analogous to the previous case but with  
next--neighbor plaquettes of the dual lattice having the same  
well--defined chirality (Fig.~\ref{twoloops}). The energy change  
$\delta E$ associated with this process is exactly that obtained for 
 $| \psi_{02} \rangle$. Any other process generated by $H_1$ has an  
associated graph of length $L(G^{ab}) > 14$, and because the amplitude  
of these processes falls in proportion to $n^{ab}_{ns} 2^{-L(G^{ab})/2}$  
($n^{ab}_{ns}$ is the number of pairs $\langle i,j \rangle$ on the same  
loop separated by a distance greater than 2), these may safely be cut  
off beyond a certain length $L_c(G^{ab})$.  
 
\subsection{Variational Wave Function} 
 
\begin{figure}[t!]
\vspace*{-0.5cm}
\hspace*{-0.5cm}
\includegraphics[angle=90,width=9cm]{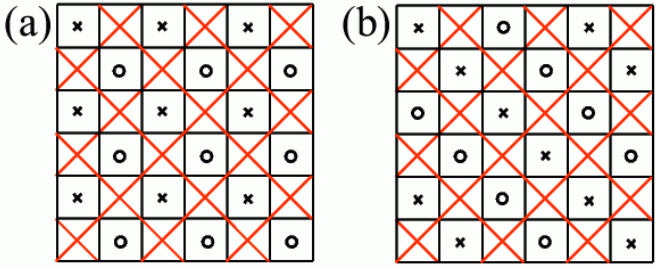}
\vspace*{-5.5cm}
\caption{ (Color online.) (a) N\'eel and (b) collinear alternation of 
uncrossed plaquettes on two sublattices. (a) Decomposition of uncrossed 
plaquettes of the checkerboard system into two sublattices $A$ (crosses) 
and $B$ (circles), which dictates the form of the variational wave 
function (\ref{var_form}) for $\Delta J > 0$; (b) decomposition into 
sublattices $C$ (crosses) and $D$ (circles) which yields the form
of the low--energy wave function for $\Delta J < 0$.}
\label{sublattices}
\end{figure}

In order to propose a wave function $|\psi \rangle$ which takes advantage  
of the three processes giving non--zero contributions for $L(G^{ab}) \leq 
14$, we introduce the bosonic operators $\Delta^{\dagger}_{\bar \alpha}$  
and $\Delta^{\;}_{\bar \alpha}$. To define these operators we classify  
the uncrossed plaquettes of the checkerboard lattice according to the  
sublattice $A$ or $B$ (Sec.~IV) of the dual square lattice  
[Fig.~\ref{sublattices}(a)]. In the six--vertex representation of the  
antiferroelectric state (Fig.~\ref{afes}), these two sublattices have  
opposite chirality, corresponding to a staggered flux phase. If the  
uncrossed plaquette ${\bar \alpha}$ has a well--defined chirality,  
the operator $\Delta^{\dagger}_{\bar \alpha}$ raises the flux from  
negative to positive if ${\bar \alpha} \in A$ and lowers it from  
positive to negative if ${\bar \alpha} \in B$,  
{\it i.e.}~$\Delta^{\dagger}_{\bar \alpha}$ changes the direction of  
arrow circulation from counterclockwise to clockwise if ${\bar \alpha}  
\in A$ and conversely if ${\bar \alpha} \in B$. For states of no  
well--defined chirality on plaquette ${\bar \alpha}$, and when  
the flux is positive (negative) for ${\bar \alpha} \in A$ ($B$),  
$\Delta^{\dagger}_{\bar \alpha}$ becomes the null operator,  
$\Delta^{\dagger}_{\bar \alpha} |\psi_a \rangle = 0$.  
 
Minimizing the energy of a variational state is equivalent to maximizing  
the number of allowed tunneling processes with a negative sign. The  
appropriate dimer background for this optimization is provided by the  
antiferroelectric state $|\psi_{\rm afe} \rangle$ of Fig.~\ref{afes};  
only for this state can the elementary process of Fig.~\ref{RK} be  
applied to every uncrossed plaquette. Moreover, the three processes which  
give finite contributions to $\delta E$ for $L(G^{ab}) \leq 14$ can be  
applied to any pair of nearest--neighbor (Fig.~\ref{graph}) and  
next--neighbor (Figs.~\ref{graph2} and \ref{twoloops}) plaquettes of  
the dual lattice when the antiferroelectric background is suitably  
populated with the local defects (bosons) created by $\Delta^{\dagger}_{\bar  
\alpha}$. For positive $\Delta J$, these processes lead to a uniform energy  
gain ($\delta E <0$) when the bosons are created with opposite phases on the 
sublattices $A$ and $B$ [Fig.~\ref{sublattices}(a)]. The considerations for  
$\Delta J < 0$ mirror those for positive $\Delta J$ with a change of the  
decomposition in two sublattices: for negative $\Delta J$ the uniform  
energy gain occurs when the bosons are created with opposite phases on  
the sublattices $C$ and $D$ [Fig.~\ref{sublattices}(b)]. Employing the  
definitions and observations introduced earlier in this section, the  
simplest wave functions for approximating the ground state in the presence  
of $H_1$ are 
\begin{eqnarray} 
| \psi_{+} \rangle & = & u^N \prod_{{\bar \alpha} \in A} (1 +   
\Delta^{\dagger}_{\bar \alpha}) \prod_{{\bar \alpha} \in B} (1 -   
\Delta^{\dagger}_{\bar \alpha}) |\psi_{\rm afe} \rangle, \nonumber \\ 
| \psi_{-} \rangle & = & v^N \prod_{{\bar \alpha} \in C} (1 +   
\Delta^{\dagger}_{\bar \alpha}) \prod_{{\bar \alpha} \in D} (1 -   
\Delta^{\dagger}_{\bar \alpha}) |\psi_{\rm afe} \rangle, 
\label{var_form} 
\end{eqnarray} 
where $|\psi_{\pm} \rangle$ correspond to $\Delta J = \pm |\Delta J|$.  
The coefficients $u$ and $v$ are determined from the normalization  
of the corresponding wave functions [Eq.~(\ref{norm})].  
Fig.~\ref{sublattices} illustrates the sublattice decompositions of  
the checkerboard geometry  for the two cases: $(A,B)$ for $\Delta J  
> 0$ ($| \psi_{+} \rangle$) and $(C,D)$ for $\Delta J < 0$ ($| \psi_{-}  
\rangle$). A similar decomposition into plaquettes $C'$, $D'$ is obtained  
by a $\pi/2$ rotation of Fig.~\ref{sublattices}(b).  
 
The wave functions $|\psi_{+} \rangle$ and $|\psi_{-} \rangle$ represent  
linear combinations including all possible two--plaquette resonance processes, 
the phase relations of which are specified by Fig.~\ref{sublattices}.  
Although these are superpositions of many states from the manifold of  
dimer coverings, the number of participating coverings remains a very  
small fraction of the total, and the selected linear combinations are  
VBC states. In the pseudospin $\tau = 1/2$ space generated by the two 
possible states of each uncrossed plaquette of $|\psi_{\rm afe} \rangle$,  
$| \psi_{+} \rangle$ and $|\psi_{-} \rangle$ are analogous to the classical,  
magnetically ordered N\'eel [${\vec q} = (\pi,\pi)$] and collinear  
[${\vec q} = (0,\pi)$ or ${\vec q} = (\pi,0)$] states polarized in  
the ${\hat x}$--direction. In this space $|\psi_{\rm afe} \rangle$  
corresponds to one of the fully polarized ferromagnetic states  
(such as $\tau^{z}_{\bar \alpha} = - 1/2$ for all ${\bar \alpha}$).  
However, $| \psi_{+} \rangle$ and $|\psi_{-} \rangle$ are not exactly  
the classical N\'eel and collinear states because two nearest--neighbor  
uncrossed plaquettes cannot be ``flipped'' independently of each other, 
{\it i.e.} 
\begin{equation} 
\Delta^{\dagger}_{\bar \alpha} \Delta^{\dagger}_{\bar \alpha'} |\psi_{\rm  
afe} \rangle = 0 
\end{equation} 
if ${\bar \alpha}$ and ${\bar \alpha'}$ are nearest--neighbor uncrossed 
plaquettes. In the pseudospin representation the two plaquettes cannot be  
simultaneously in the $\tau^z = 1/2$ state, as a result of which these  
states are projected out of the superpositions corresponding to the  
classical N\'eel and collinear states to obtain $| \psi_{+} \rangle$  
and $|\psi_{-} \rangle$. We stress that the symmetries of $|\psi_{+}  
\rangle$ and $|\psi_{-} \rangle$ for $\Delta J = \pm |\Delta J|$  
are quite different. 
  
The energies of the wave functions $| \psi_{+} \rangle$ and $|\psi_{-}  
\rangle$ obtained from Eq.~(\ref{general_energy}) are  
\begin{eqnarray} 
E^+ = \langle \psi_{+}| H_1 | \psi_{+} \rangle & = & E_d - {\textstyle  
\frac{3}{16}} u^2 N  \Delta J \nonumber \\ 
E^- = \langle \psi_{-}| H_1 | \psi_{-} \rangle & = & E_d + {\textstyle  
\frac{3}{16}} v^2 N  \Delta J . 
\label{vare} 
\end{eqnarray} 
The three processes represented in Figs.~\ref{graph}, \ref{graph2}, and 
\ref{twoloops} give an extensive negative contribution to $E^{\pm}$.  
The lack of frustration among these different processes indicates that 
$| \psi_{+} \rangle$ and $|\psi_{-} \rangle$ are representative of  
the VBC orderings which are stabilized respectively for $\Delta J > 0$  
and $\Delta J < 0$. We draw particular attention to the fact that in  
these valence--bond orderings the four dimers around each uncrossed  
plaquette resonate between the two configurations shown in Fig.~\ref{RK}.  
The relative phase is positive on one sublattice ($A$ for $\Delta J > 0$ 
and $C$ for $\Delta J < 0$) and negative for the other ($B$ for $\Delta J  
> 0$ and $D$ for $\Delta J < 0$). Note that the RK process would lead to 
a different ground state in which this relative phase is the same for 
all the plaquettes of the dual lattice. In the pseudospin basis this 
corresponds to the ferromagnetic state [${\vec q} = (0,0)$] with  
polarization in the ${\hat x}$--direction for $\Delta J < 0 $ and  
the $-{\hat x}$--direction for $\Delta J > 0 $. 
 
With regard to possible four--spin perturbation interactions, we remark  
that the dominant type of such a perturbation would be that contained  
in the second term of Eq.~(\ref{esh}). However, the effect of altering  
$K$ is trivially identical to the perturbation of  
Eq.~(\ref{Heisenberg_perturbation}), and the results derived throughout  
this section may be applied directly to Eq.~(\ref{esh}) with a sign  
inversion of $\Delta K \equiv (K - K_{c})$ relative to $\Delta J$. 
 
We now summarize the extension of our results to the pyrochlore lattice.  
Here any given dimer covering leads to precisely the same, uniform energy  
shift (\ref{uniform}) when a supplementary nearest--neighbor Heisenberg  
interaction is inserted. When superpositions of dimer coverings are  
considered, the elementary graph (or analog of the RK process)  
circumscribes a hexagon formed by six neighboring tetrahedra. This  
elementary hexagonal loop replaces the sides of the uncrossed plaquettes  
on the checkerboard lattice. As for the checkerboard, the graph  
corresponding to the elementary process is simple, implying that this  
process is not generated by the perturbation $H_1$. Once again it is  
necessary to consider larger loops to find the valence--bond ordering  
stabilized by the perturbation. 
 
\begin{figure}[t!]
\vspace*{-0.5cm}
\hspace*{-1.2cm}
\includegraphics[angle=90,width=10cm]{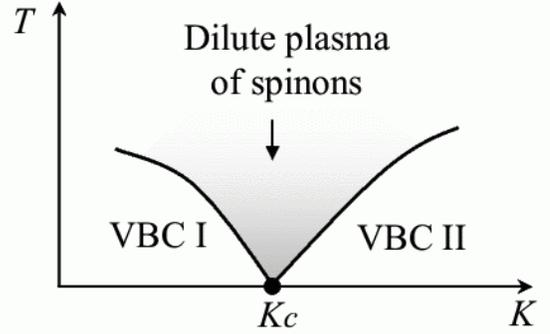}
\vspace{-4.7cm}
\caption{ Low--temperature phase diagram around the Klein point for
checkerboard and pyrochlore spin systems.}
\label{pdiag}
\end{figure}

As stated above, the local zero--divergence constraint defining the  
low--energy subspace around the Klein point implies that the low--energy  
theory induced by a general perturbation is always a compact U(1) gauge  
theory. Following the argument of Polyakov, the phase stabilized by a  
general perturbation is always confining at $T = 0$. For the spin models  
we consider, this confining phase consists of a VBC configuration.  
The perturbation $H_1$ treated in this subsection is a particular example of  
this general observation, and also perhaps the most physically relevant. The  
effect of a general perturbation in the space of interaction parameters is  
illustrated in the schematic phase diagram of Fig.~\ref{pdiag}.  
 
We conclude this section with a summary: we have introduced a  
loop--based approach to account for the physical processes in  
short--ranged quantum dimer systems with only local interactions,  
which is powerful and general. We have used this procedure to compute  
energies and variational wave functions in the non--orthogonal basis  
of dimer coverings on the checkerboard lattice. These considerations  
include all of the physically relevant perturbations of the Klein  
Hamiltonian, as discussed in Sec.~II, either explicitly or by direct  
analogy. We conclude that the general case is one in which VBC phases,  
which are gapped and confined, are stable away from the Klein point. 
 
\section{Exact criticality in a finite--temperature region} 
\label{exact-criticality} 

At issue for the physical properties of a system is rather less the 
behavior at $T = 0$ than that at finite temperature. In the preceding 
section we devised a systematic, purely topological, diagrammatic method 
to compute the energies of all variational wave functions in the dimer 
basis. This procedure is exact for the degenerate manifold of basis 
states, including most importantly when this is non--orthogonal, and 
as shown in Sec.~V presents a tractable problem at zero temperature. At 
finite temperatures the need to evaluate brackets between all possible 
states in the Hilbert space presents a more serious problem. In this 
section we discuss the extent to which progress is possible in this 
regime, by first illustrating how to establish the existence of a 
finite--temperature region of critical behavior for general pertubations, 
and then consider a simple, anisotropic perturbation to (a) demonstrate 
explicitly the existence of this critical region and (b) derive its 
approximate phase boundaries. In the remainder of this section we set 
$k_{\rm B} = 1$ and for rigor assume implicitly that $J/T \to \infty$. 
$J$ is the gap scale separating the ground state energies from 
the lowest excited states. The presence of a spectral
gap in a Klein model was shown in \cite{Raman05}.


\subsection{Finite--temperature criticality}
\label{finite}

Consider a general perturbation $H^{\prime}$ which augments the 
Klein--point Hamiltonian to give 
\begin{equation}
H_{tot} = H + H^{\prime};
\label{lambda}
\end{equation}
Eq.~(\ref{Heisenberg_perturbation}) provides one specific example. We 
employ the notation $\epsilon$ (Sec.~IV) to denote the small energy scale 
of $H^{\prime}$. In what follows, we prove that unless the ground states 
of $H_{tot}$ are ordered, in the sense that the different ground states 
are distinguishable by local order parameters, then a region of 
finite--temperature criticality must occur if the perturbation is
not linear in an order parameter. 

We begin by stating that the system displays critical correlations (i) 
when $\epsilon = 0$ at any finite temperature and (ii) at $T/\epsilon \to\infty$. 
We also note that (iii) any local order parameter vanishes for any fixed $\epsilon$ 
at sufficiently high temperatures $T > T_{c}(\epsilon)$. Finally we
assume (iv) that at sufficiently high temperature the two--point
correlator of an order parameter decreases monotonically in temperature
at long enough distances. While (i), from Sec.~IV, 
and (iii) are self--evident, the demonstration of statement (ii) represents 
the core of the proof: because critical correlations appear in 
the high--temperature limit, (iii) and (iv) imply that 
correlations at any finite but high enough temperature cannot be weaker
than algebraic. Thus the system exhibits critical correlations at
sufficiently high but finite temperature.

The derivation of statement (ii) proceeds from the definition of the 
relevant order parameter. By analogy with the classical dimer polarization 
operators presented in Sec.~IIIC, we define the corresponding quantum 
operators. For a given plaquette A formed by the four sites $(a,b,c,d)$ 
appearing in counterclockwise order, the polarization is given by  
\begin{eqnarray}
\vec{P}_{A} & \equiv & P_{A;x} \hat{e}_{x} + P_{A;y} \hat{e}_{y}
+ P_{A;z} \hat{e}_{z} \nonumber
\\ & = & \Big[ (\vec{S}_{a} \cdot \vec{S}_{d} - \vec{S}_{b} 
\cdot \vec{S}_{c}) \hat{e}_{x} + (\vec{S}_{a} \cdot
\vec{S}_{b} - \vec{S}_{c} \cdot \vec{S}_{d} ) \hat{e}_{y} \nonumber
\\ & & \;\;\;\;\;\; +
 (\vec{S}_{a} \cdot \vec{S}_{c} - \vec{S}_{b} \cdot \vec{S}_{d}) 
\hat{e}_{z} \Big]. 
\end{eqnarray}
The dimer--dimer correlation function between two crossed plaquettes 
$A$ and $B$ is then given, under the assumption that all states in the Klein 
manifold are linearly independent \cite{Chayes89}, and at finite 
temperatures $T \gg \epsilon$, by the expectation value 
\begin{eqnarray}
\chi_{AB} & = & \langle \vec{P}_{A} \cdot \vec{P}_{B} \rangle_{\epsilon, 
T > 0} \nonumber
\\ & = & \frac{Tr\Big[ e^{-\beta H} \vec{P}_{A} \cdot \vec{P}_{B}  \Big]}
{Tr \Big[e^{-\beta H}\Big]} \nonumber 
\\ \lim_{T/\epsilon \to \infty} \chi_{AB} & = & \frac{1}{N_{g}} \sum_{|\psi 
\rangle} \langle \psi| \vec{P}_{A} \cdot \vec{P}_{B} |\psi \rangle \nonumber
\\ & \propto & ~ |r_{AB}|^{-d},
\label{explain_ortho}
\end{eqnarray}
with $|r_{AB}|$ the separation between the centers of the plaquettes $A$ 
and $B$, $d$ the spatial dimensionality, $N_{g}$ the number of states in 
the ground--state manifold (Sec.~IIIA) and the sum taken over a complete 
set of orthonormal states $\{| \psi \rangle\}$. At the Klein point, the 
correlation function $\chi_{AB}$ is algebraic, as shown in Sec.~IV. 

The essential property underlying the physics of the finite--temperature 
derivation is that the value of $\chi_{AB}$ is the same for all temperatures 
when the magnitude $\epsilon$ of the external perturbation vanishes. This 
is a consequence of the fact that all states within Klein--point manifold 
carry equal weight, whence $\chi_{AB}$ is given by the last two lines of 
Eq.~(\ref{explain_ortho}) and the system is manifestly critical. As the 
Klein--point limit is approached for $T/\epsilon \to \infty$, this average 
becomes equivalent to a sum over all states spanning the Klein--point 
Hilbert space and its value is precisely the same as for the $T = 0^{+}$ 
case when $\epsilon$ is taken to zero. Here the probability of any state 
$|\psi \rangle$ as given by the finite--temperature density matrix is 
$\exp[-\beta E_{\psi}]/Z = 1/N_{g}$. For $T/\epsilon \to \infty$ the 
correlation function $\chi_{AB}$ then takes the form specified by 
Eq.~(\ref{explain_ortho}), decaying algebraically to $0$ as stated in 
(ii) above.

Although it is not relevant to the above derivation, we remark briefly 
that topological rules similar to those of Sec.~V may be devised for the 
computation of the dimer--dimer correlation function. As a consequence of the 
two scalar products appearing in each term of $\chi_{AB}$, two permutation 
operations can be performed on the loop formed by two states (as opposed 
to the single operation discussed in Sec.~V). A similar set of topological 
rules for the computation of $\chi_{AB}$ will be provided elsewhere. 

The use of statements (ii) and (iii) for arbitrary pertubations is 
assisted by considering the form of general order parameters. Let 
$\{|g_{\alpha} \rangle\}$ denote the ground states of $H_{tot}$ and 
$\hat{O}$ the order parameter by which they are distinguished. In a 
system which displays local order, $\hat{O}_{\bf r}$ is an operator 
which depends on a finite number of fields in a sphere (or disk) of 
finite radius $R$ which surrounds a given point ${\bf r}$). A general 
algorithm for the construction of the order parameter by examining 
reduced density matrices is given in Ref.~\cite{FMO}. 

At $T = 0$ within any ground state, $\langle \hat{O}_{\bf r} \hat{O}_{\bf r'} 
\rangle \to |\langle \hat{O}_{\bf r} \rangle|^{2} \equiv m^{2} \neq 0$ for 
separations $|{\bf r} - {\bf r'}| \to \infty$. Following 
Eq.~(\ref{explain_ortho}), 
\begin{equation}
\langle \hat{O}_{\bf r} \rangle = \frac{\sum_{\psi,|\phi_{i} \rangle} \langle \psi| \phi_{1} 
\rangle M_{12} \langle \phi_{2}| \hat{O}_{\bf r} | \phi_{3}  \rangle M_{34} 
\langle \phi_{4} | \psi \rangle e^{-\beta E_{\psi}}}
{\sum_{\psi} e^{-\beta E_{\psi}}},
\label{average_explicit}
\end{equation}
where $\{|\phi_{i} \rangle\}$ denotes a set of pure dimer states and 
$\{|\psi \rangle \}$ a complete set of orthonormal states spanning the 
Klein--point basis. Equation (\ref{average_explicit}) makes use of the 
overcompleteness relation
\begin{eqnarray}
\sum_{\phi_{1}, \phi_{2}} |\phi_{1} \rangle M_{12} \langle  \phi_{2}|
= 1_{\rm Klein},
\label{KP}
\end{eqnarray}
with $1_{\rm Klein} = \prod_{\boxtimes} (1-{\cal{P}}^{\boxtimes})$ the 
unit operator in the Klein--point subspace [and ${\cal{P}}^{\boxtimes}$ 
the projection operator onto the subspace of net spin $S_{\boxtimes} = 2$ 
(Sec.~II)]; Eq.~(\ref{KP}) expresses the fact that the dimer states completely 
exhaust the Klein--point subspace, as proven in Ref.~\cite{nussinov_June06}.
As emlpoyed in \cite{Chayes89}, all Klein model ground states
can be expressed as a sum of projection operators on these ground
states which decompose them into 
(Young tableaux) sectors which are well defined under the permutation 
of pair sites. The presence of a singlet dimer
between two sites enforces antisymmetry 
amongst the pertinent lattice sites. As evident 
in expressing the ground states in terms of the irreducible representations
of the permutation group, the most general operations allowed within 
this basis amount to permutations. \cite{explain_Young}
As the most general permutation can be written
as a product of pairwise permutations and 
as $\vec{S}_{i} \cdot \vec{S}_{j} = \frac{1}{2} (P_{ij} - \frac{1}{2})$, 
where $P_{ij}$ is the operator permuting sites $i$ and $j$,
the most general permutation is
a functional of scalar spin 
products. For an operator $\hat{O}_{\bf r}$ which is 
a functional of spin products of 
the form $\vec{S}_{a} \cdot \vec{S}_{b}$, 
the correlation function $\langle 
\hat{O}_{\bf r} \hat{O}_{\bf r'} \rangle$ is equivalent to a correlation 
function between dimer products. From the analogy to a dipolar system 
(Sec.~III), correlation functions containing more scalar spin products 
decay algebraically according to the strength of the Coulomb interaction 
between more complicated dipole configurations. Moments of higher order
than dipolar are important for those dimers which are close to each other 
but far from all other dimers. All correlation functions thus decay 
algebraically with the dimer--dimer separation. The analysis based on 
Eq.~(\ref{explain_ortho}) may now be extended to more general correlation 
functions which involve the order--parameter fields $\hat{O}_{\bf r}$. 

Returning to statement (iii), as in our earlier considerations for 
the polarization $\vec{P}$ this is based on the result that any 
finite--temperature expectation value $\langle \hat{O}_{\bf r} 
\hat{O}_{\bf r'} \rangle$ will decrease monotonically with increasing 
temperature, and asserts that when the expectation value of the order 
parameter vanishes, $\langle \hat{O}_{\bf r} \rangle = 0$, the 
corresponding energy scale sets the transition temperature $T_{c}$ 
[(iii)]. This quite universal property is reviewed briefly in 
Ref.~\cite{entropy-energy}. We reiterate [(ii)] the fact that in 
the high--temperature limit, where correlations are weakest, the 
correlation function $\langle \hat{O}_{\bf r} \hat{O}_{\bf r'} \rangle$ 
retains a decay which is only algebraic in the separation $|{\bf r} - 
{\bf r'}|$. From these results it is clear that there is a 
finite--temperature region surrounding the ``Klein line'' $(\epsilon 
= 0, T > 0)$, and specified explicitly by $T > T_{c}(\epsilon)$, where 
critical correlations appear.

We conclude the general discussion by noting that for a perturbation which 
is linear in a general order parameter, $H' = - \epsilon  \sum_{A} Q_{A}$, 
it is possible that the two--point correlation function takes a finite 
value, $\langle Q_{A} Q_{B} \rangle \to m^{2} \neq 0$ as $|r_{AB}| \to 
\infty$, at any finite $T$. This type of perturbation may then yield 
correlations which are algebraic only in the limit $T \to \infty$ 
\cite{exception}. In the next subsection we consider a particular 
perturbation whose effect can be approximated by the anisotropic 
six--vertex model to show that the dimer correlations may also remain 
critical up to a threshold value of the perturbation. 

\subsection{Anisotropic perturbation}

In this subsection we use a specific anisotropic perturbation to 
investigate the nature of the phase diagram. We will illustrate 
explicitly that the phase diagram contains a finite--temperature 
critical region, exemplifying the general concepts of the preceding 
subsection, and then derive its approximate phase boundaries.

We consider a perturbation of the form  
\begin{equation} 
H_{d} = J_{d} \sum_{\langle \langle ij \rangle \rangle} {\vec S}_{i}  
\cdot {\vec S}_{j}, 
\label{H_d} 
\end{equation} 
where $\langle \langle ij \rangle \rangle$ denotes all diagonal pairs on  
the same crossed plaquette. This type of interaction emulates the weakening  
(for $J_{d} < 0$) of exchange interactions between neighboring diagonal  
sites on a checkerboard plaquette relative to those of horizontal or  
vertical spin pairs, or a lifting (to tetragonal) of the cubic symmetry  
in a pyrochlore tetrahedron. Focusing for specificity on the checkerboard  
geometry, for $H_{tot} = H + H_{d}$ it is clear that in the  
regime $J_{d},T \ll J$, the effect of $H_{d}$ is to favor  
certain dimer, or equivalently six--vertex, configurations over others,  
and in particular that for $J_{d} > 0$ the diagonal contribution, $\langle  
\psi_a| H_d | \psi_a \rangle$, favors singlet formation on the diagonal  
bonds of all plaquettes. Because off--diagonal brackets are significantly  
smaller than diagonal contributions, these may be neglected. The favored 
ground states in this case are precisely the two antiferroelectric states. 
Similarly, for $J_{d} < 0$ the single dimer on each crossed plaquette lies 
preferentially on the vertical or horizontal bonds. Because only the diagonal 
contributions $\langle \psi_a| H_d | \psi_a \rangle$ are considered, the 
ground--state degeneracy for $J_{d} > 0$ remains exponentially large:  
the system supports that subset of all allowed ground states which  
corresponds to the ground--state manifold of the Klein model on the  
square lattice, the size of which scales with $2^{1 + \sqrt{N}}$, 
where $N$ is the number of lattice sites \cite{rbt}. We remind the 
reader that the states in this restricted set of dimer coverings, in 
which precisely one dimer is present on every square--lattice plaquette, 
are strictly orthogonal to each other in the thermodynamic limit. The 
general case has a very much larger manifold of ground states, the 
classification of which proceeds by the line representation presented 
in Sec.~III. 

\subsubsection{Critical region}

The illustration that the phase diagram in the presence of this perturbation 
contains a finite--temperature critical regime is based on the exact limits 
examined in Subsec.~(\ref{finite}) and constitutes an explicit implementation
of this discussion. We begin by noting, as before, that when $J_{d} = 0$ the 
system remains at the Klein point for all temperatures, displaying critical 
correlations, and the analysis of Sec.~IV is applicable. The susceptibilities 
$\chi^{aa}_{AB} = \langle P_{A;a} P_{B;a} \rangle$ are a direct probe of dimer 
correlations: at $J_{d} = 0$ the correlation function $\chi_{AB}$ is critical 
for all $T > 0$.

With reference to the considerations below Eq.~(\ref{H_d}), for $J_{d} > 0$, 
$\chi^{zz}_{AB}$ increases in magnitude relative to its Klein--point value 
for diagonal dimers only. Thus, {\it (i)} the decay of $\chi$ with distance 
is no faster than algebraic. However, no VBC order is possible at high 
temperatures $T \gg J_{d}$, and thus {\it (ii)} $\chi \to 0$ as the 
plaquette separation $|r| \to \infty$. Conditions {\it (i)} and {\it (ii)} 
taken together imply that for $T \gg J_{d}$ the system exhibits algebraic 
correlations. Similar considerations apply when $J_{d} < 0$, where 
$[|\chi^{xx}_{AB}| + |\chi^{yy}_{AB}|]$ increases relative to its 
Klein--point value, corresponding to correlations of horizontal or 
vertical dimers. We thus conclude that for $T \gg |J_{d}|$ the system 
is critical. These statements are completely general, and provide one 
explanation for why it is that the six--vertex model always has critical, 
and only critical, correlations at sufficiently high 
temperatures. \cite{marginalconfinement}

\subsubsection{Finite--temperature criticality by approximate
mapping to an anisotropic six--vertex model}

We now derive the approximate forms of the phase boundaries of the 
critical region. This calculation is not exact for the reasons stated 
at the start of this section, and is included to provide a qualitative 
indication of the nature of the phase diagram. We begin with an explicit 
statement of the approximations employed in this derivation. {\it (i)} 
In contrast to Sec.~V, for this finite--temperature derivation the 
non--orthogonality of the different dimer states is neglected. At 
the Klein point ($J_{d} = 0$) all dimer configurations carry equal 
weight and their orthogonality is irrelevant, a fact exploited in 
Sec.~V. All of the properties of the system are a consequence of the 
entropic considerations presented in Sec.~IV. However, away from the 
Klein point the configurations no longer have equal probability 
(in the example of Eq.~(\ref{H_d}), states with diagonal dimer 
configurations are favored when $J_{d} > 0$, and suppressed when 
$J_{d} < 0$). {\it (ii)} Off--diagonal contributions in the full 
Hamiltonian, {\it i.e.}~terms of the form $\langle \psi_{a} |H| 
\psi_{b} \rangle$ with $a \neq b$, are neglected.   

When mapped to the six--vertex representation, provided that the 
conditions $J_{d}, T \ll J$ are satisfied, the anisotropic perturbation 
of Eq.~(\ref{H_d}) is reflected in a change in the relative weights of 
the six--vertex configurations (which at low temperatures are the only 
states with non--vanishing weights). At $T = 0$, for $J_{d} > 0$ these 
nonuniform weights are represented by vanishing energies for four of the 
vertex types while the other two contribute an equal and finite energy, 
and conversely for $J_{d} < 0$. 

\begin{figure}[t!]
\vspace*{-0.7cm}
\hspace*{0cm}
\includegraphics[angle=90,width=11cm]{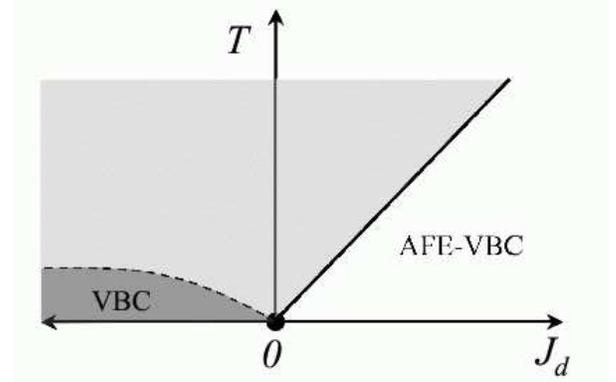}
\vspace*{-5.8cm}
\caption{Approximate phase diagram for Klein--point Hamiltonian with 
diagonal exchange perturbation $J_d$. Neglecting both non--orthogonality
and off--diagonal contributions, an antiferroelectric valence bond crystal 
(AFE--VBC) ordering is stabilized below $T = 3 J_d/ (8 \ln{2})$ for $J_d > 0$. 
The system is critical at all points within the region shaded in light grey.
In the absence of off--diagonal processes, it remains critical to zero 
temperature for $J_{d} < 0$; the qualitative effect of these processes 
is to remove the low--temperature critical behavior by stabilizing a VBC 
ordering in the region shaded dark grey. While the phase boundaries shown 
above are approximate, the existence of a finite--temperature critical region 
is not (Sec.~VIA).}
\label{JD}
\end{figure}

With reference to Fig.~\ref{rules}, we choose the energies of the  
six--vertex configurations displayed to be $\epsilon_{i=1,2,3,4} = 0$  
and $\epsilon_{5} = \epsilon_{6} = - {\textstyle \frac{3}{4}} J_{d}$  
in the presence of the perturbation. The correlations and thermodynamics  
of this model are best analyzed through the fugacities $a = \exp[- \beta  
\epsilon_{1}] = \exp[- \beta \epsilon_{2}], ~b = \exp[ -\beta \epsilon_{3}]  
 = \exp[ -\beta \epsilon_{4}]$, $c = \exp[-\beta \epsilon_{5}] = \exp[-\beta  
\epsilon_{6}]$, where $\beta$ is the inverse temperature. The fugacities 
determine a parameter $\Delta$ whose value determines in turn the phase 
in which the six--vertex model lies,  
\begin{equation} 
\Delta = \frac{a^{2} + b^{2} - c^{2}}{2ab}. 
\label{delta_val} 
\end{equation} 
Following Lieb's original solution of the six--vertex model  
\cite{Lieb,Baxter}, all models with $|\Delta| < 1$ are in the disordered 
phase. Such small--$|\Delta|$ systems, including the ice model relevant  
for the isotropic checkerboard lattice where $a = b = c = 1$, are in the  
disordered phase and exhibit power--law correlations. By contrast, for  
$\Delta > 1$ the system adopts an ordered configuration where all arrows  
have a definite chirality (such as up and to the right or down and left).  
In the system with perturbation $J_d$, $a = b = 1$ while $c = 
\exp[\frac{3}{4} \beta J_{d}]$, and one finds that if  
\begin{equation} 
T > \frac{3}{8 \ln 2} J_{d}, 
\label{T_6V} 
\end{equation} 
then $|\Delta| < 1$, indicating that the system is in its disordered phase.  
In this case we find power--law correlations among dimer pairs and the  
system is trivially critical for all non--zero temperatures when $J_{d}  
< 0$, or critical only at sufficiently high temperatures [whose lower  
bound is given by Eq.~(\ref{T_6V})] whenever $J_{d} > 0$. When $T = 0$,  
one finds that for all $J_{d} < 0$ the system lies precisely on the 
boundary  between the ordered and the critical phase $(\Delta = 1)$. 
Equation (\ref{T_6V}) defines a finite region within the $(J_{d},T)$ 
plane (Fig.~\ref{JD}) in which the system is critical.  

The above results are applicable under the conditions that all relevant 
states lie within the Klein basis manifold, {\it i.e.}~$T \ll J$, that 
the effects of non--orthogonality are neglected, and that only the 
diagonal corrections induced by $H_d$ are considered. Although these 
are small, off--diagonal contributions of the type $\langle \psi_a| H_d 
|\psi_b \rangle$, analogous to those computed in Sec.~V, are relevant 
in removing the "marginally critical" $T = 0$ correlations obtained for 
$J_{d} < 0$, and stabilize once again a particular valence--bond ordering 
pattern of the types considered in detail in Sec.~V. In this situation the  
marginally critical line may be moved to finite temperatures, separating  
the low--$T$ ordered phase from a finite--$T$ critical regime. In summary, 
it is possible to demonstrate rigorously that specific types of perturbation 
around the Klein point lead to an exotic, exactly critical regime. 
Quantitative calculations of the properites of this regime are not 
exact, but approximations of the kind applied here may be used for 
the qualitative determination of quantities such as the boundaries 
of this phase.

\section{Discussion and Conclusions}  

The pyrochlore (A$_2$B$_2$O$_7$) and spinel (AB$_2$O$_4$) structures 
are very common among magnetic oxides, occurring for a wide variety  
of transition and rare--earth metals as both the A and B ions and  
thereby offering a range of valence states and magnetic moments $S$  
in a pyrochlore lattice of interacting spins. Because the orbital  
contribution to the magnetic moment of transition metal ions in the  
3$d$ series is quenched by the crystal--field splitting, the spin--orbit  
coupling has only a perturbative effect and the relevant physical spin  
models are effectively SU(2)--invariant. While the link to a physical  
pyrochlore or spinel system satisfying all the necessary criteria remains  
to be found (Sec.~II), we have shown here that for these structures a  
class of SU(2)--invariant $S = 1/2$ models of this type, whose local  
interactions emerge directly from the simplest Hubbard Hamiltonian,  
gives rise to spinon excitations which propagate in the full lattice.  
 
At the root of the exotic behavior exhibited by this type of system is  
the massive (extensive) degeneracy of the ground--state manifold in the  
vicinity of the Klein critical point. We have gained physical insight  
into the multiple essential aspects of this problem by exploiting exact  
mappings between the spin system under consideration and other physical  
systems \cite{Batista04}. In particular, because the ground--state sector  
of the $S = 1/2$ Klein models on the pyrochlore lattice obeys ice  
rules of the type discussed in Sec.~III, it is possible to exploit  
analogies with six--vertex models \cite{Baxter}, string gases and U(1)  
gauge magnets to make a number of powerful qualitative and quantitative  
statements. The local constraint of zero divergence imposed by the ice  
rules implies that the low--energy sector is described by an effective  
U(1) gauge theory in the neighborhood of the Klein point. This theory  
is, however, not necessarily the minimal U(1) gauge magnet, whose terms  
involve only the smallest loops, which is usually invoked in these cases  
\cite{ring_exchange1}. The variational approach which we introduce and employ  
in Sec.~V indicates that in some cases, as a result of the non--orthogonality  
of the singlet dimer coverings, processes involving larger loops are  
relevant for realistic physical perturbations.  
 
Although the spinons are deconfined excitations only at a single point  
(the Klein point) of the quantum phase diagram ($T = 0$), this deconfinement  
exists over a finite region of the phase diagram due to thermal fluctuations  
which give rise to a dilute plasma phase of spinons.  
The spinon--spinon correlations in this region are dominated by the $T  
\rightarrow 0^+$ Klein critical point, whose critical fluctuations are of  
entropic origin. The primary characteristic of the Klein point is the  
stabilization of a low--energy manifold of states satisfying a local  
constraint of zero divergence, or equivalently the ice rules \cite{fowler, 
Pauling}, and the removal of any low energy scale. As a consequence of both  
conditions, the system becomes critical due to entropic fluctuations induced 
even at infinitesimal temperatures.  
 
These features are directly relevant for the potential observation  
of spinons in two-- or three--dimensional systems. In the case of the  
pyrochlore lattice models considered here, the deconfined phase is expected  
to appear near a transition between two valence--bond crystals. Similar  
considerations were applied to the $S = 1/2$ model studied in  
Ref.~\cite{rbt}, the difference in that case being that the spinons  
propagate along one--dimensional paths because the dimensional reduction  
at the corresponding Klein point is incomplete.   
 
It is also interesting to note that an exact deconfined quantum  
critical point (QCP) \cite{senthil04} can be induced by adding a  
Zeeman term to the Klein Hamiltonian of Eq.~(\ref{ho}). The QCP  
is induced at the critical magnetic field $B_c$ which closes the  
singlet--triplet spin gap. The excited two--spinon deconfined state 
becomes the ground state at $B_c$. For higher values of the field  
($B > B_c$), the concentration of spinons becomes finite and the  
degeneracy between the underlying singlet dimer coverings is lifted.  
Consequently, spinon--spinon confinement is restored and the resulting  
triplet pairs are expected to condense, giving rise to an XY--type  
antiferromagnet (XY--AF) in the plane perpendicular to the applied field.  
 
\begin{figure}[t!]
\vspace*{-0.6cm}
\hspace*{-1.0cm}
\includegraphics[angle=90,width=9cm]{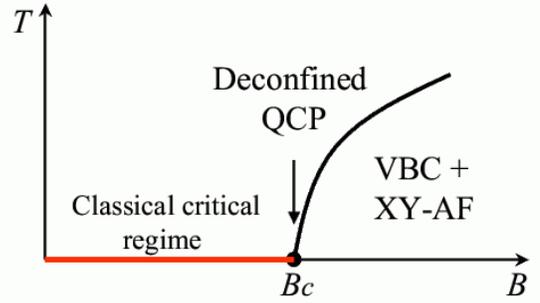}
\vspace{-4.3cm}
\caption{ (Color online.) Low--temperature phase diagram for the Klein 
Hamiltonian in the presence of a Zeeman term (applied magnetic field $B$) 
for the checkerboard and pyrochlore spin systems.}
\label{pdiagb}
\end{figure} 
 
In the presence of a magnetic field, the system at the Klein point  
displays a true deconfined QCP between a line of classical critical  
points in the region $B < B_c$ and a magnetically ordered region at  
$B > B_c$ with two coexisting order parameters. A schematic phase  
diagram illustrating the finite extent of the deconfined regime is  
shown in Fig.~\ref{pdiagb}. Very different and much more conventional  
behavior is obtained away from the Klein point, where the ground--state  
degeneracy is lifted. Here the region $B < B_c$ has a VBC ground state  
with gapped magnon excitations, separated by a conventional QCP from an 
 ordered phase with spin--wave excitations.  
 
The ideas and concepts connected with deconfinement at quantum critical  
points have formed an extensive recent literature best summarized in  
Ref.~\cite{senthil04}. As noted in Sec.~I, many of these studies depart  
from effective U(1) gauge--theoretical treatments rather than microscopic  
models, and it is not yet clear that specific systems exist which realize  
the desired deconfinement properties. 
The Klein point in certain pyrochlore lattice models was considered  
briefly in Ref.~\cite{Raman05}, although these authors did not dwell on  
either the detailed physical properties of such a point or on their origin.  
We have found that the mapping used in this study to a quantum dimer model  
with only RK interactions is not in fact justified for a Heisenberg  
antiferromagnet in a pyrochlore geometry. 
The scenario of the ``constrained entropic critical point'' introduced  
recently in Ref.~\cite{rcastelnovo} contains certain parallels to the  
physics of the Klein point with regard to high--temperature constraints. 
However, this is said to be a ``top--down'' construction based on different  
energy scales, some of which are present in a number of the effective models  
considered, and does not contain specific properties emerging from  
a microscopic Hamiltonian for all temperature regimes. 
 
We conclude this discussion with a speculation. The exponential ground--state  
degeneracy [Eqs.~(\ref{pauling.},\ref{elliot.})] and resulting extensive  
configurational entropy which we have found suggest, but certainly do not  
mandate, that glassy spin dynamics may occur naturally in pyrochlores. 
We stress that these systems are uniform and disorder--free  
but frustrated, whence this suggestion would reinforce the widely held  
belief that frustration rather than disorder is the fundamental requirement  
for the dynamical properties of structural glasses. The basic premise  
of the Vogel--Fulcher (VF) form of glassy dynamics is that the relaxation  
times appear to diverge at a temperature which correlates well with the  
intercept ($T_{0}$) of the extrapolated entropy of the super--cooled liquid  
to that of the solid. According to a prevalent line of reasoning \cite{kirk},  
as the system is cooled the configurational entropy first becomes extensive  
at $T_{A}$, the onset temperature of multiple local free--energy minima;  
at a lower temperature, $T_{0}$, these minima become stable and the  
configurational entropy vanishes. If an exponentially large number of 
metastable states is found for $T_{A} > T > T_{0}$, one may invoke the  
analysis of entropic droplets \cite{kirk} to obtain the characteristic  
free--energy barrier height 
\begin{equation} 
\Delta E  \propto (T S_{c})^{-1}. 
\end{equation} 
Linearizing the extensive configurational entropy, 
\begin{equation} 
S_{c}(T \to T_{0}^{+}) \sim V (T/T_{0}-1), 
\end{equation} 
leads to VF dynamics \cite{kirk} with relaxation times  
\begin{equation} 
\tau \sim \exp[D T_{0}/(T - T_{0})]. 
\label{VFT} 
\end{equation} 
Here $T_{0}$ a temperature scale close to the Kauzmann temperature  
\cite{Kauzmann} at which an ``ideal glass transition'' would occur, and  
where the extrapolated entropy of the assumed liquid undergoes a crisis.  
We stress that this very general derivation of a systematic definition  
for glassy behavior, by consideration of free--energy barriers \cite{kirk},  
does not require that the different ground states be linked only by  
infinite--length processes. (However, if only the spatially longest  
processes were operational this would indeed lead to very robust,  
exponentially slow quantum dynamics, as first noted by Chamon and  
coworkers \cite{Chamon}.) For the pyrochlore system we have shown that  
there exist finite--length tunneling processes linking different ground  
states within the highly degenerate low--energy sector, and thus 
suggesting an obvious candidate system for this type of glass. We  
comment that the physical processes indicated here, namely of glassy 
spin dynamics arising from frustration in a periodic lattice, are similar 
to early ideas of the ``topological spin glass'' of Ref.~\cite{coleman}.  
Empirically, glassy dynamics are strongly indicated in a number of frustrated  
magnetic systems, of which we quote only a selective list: evidence of phases  
with certain glassy characteristics has been reported in some pyrochlores,  
including Y$_{2}$Mo$_{2}$O$_{7}$ \cite{Gardner,Ladieu}, in the stacked kagome  
layer system SrCr$_{8.6}$Ga$_{3.4}$O$_{19}$ \cite{Ladieu}, and in the kagome  
bilayer compound Ba$_{2}$Sn$_{2}$ZnCr$_{6.8}$Ga$_{3.2}$O$_{22}$ \cite{Bono}. 
Further indications for glassy behavior are found in some quasi--triangular  
antiferromagnets, including NiGa$_{2}$S$_{4}$ \cite{Nakatsuji}. 
 
In conclusion, we have performed a detailed analysis of a physically  
motivated quantum spin model with only near--neighbor interactions on  
the two-- and three--dimensional pyrochlore lattices. This reveals a  
wealth of exotic behavior which can be traced to the extensive degeneracy  
of the system at the Klein point and thus to a complete dimensional  
reduction. The complete absence of quantum fluctuations in the ground--state  
manifold at the Klein point leads to a new type of classical critical point, 
with exactly known critical behavior, which we stress emerges from a highly  
frustrated, microscopic quantum spin model. 
 
The critical correlations at the Klein point are driven by entropic  
(thermal) fluctuations, leading to an effective plasma phase with  
Coulombic interactions. These are not confining, whence the elementary  
excitations are deconfined spinons which propagate freely in all  
directions (a further manifestation of complete dimensional reduction).  
This high--dimensional fractionalization would also be manifest as a  
spin--charge separation in the dilute limit of added holes if the 
carrier hopping were much smaller than the magnetic energy scale $J$.  
Physically relevant perturbations away from the Klein point lead in  
general to confined phases of static valence--bond order with spin gaps  
to $S = 1$ excitations. However, at finite temperatures in the vicinity  
of the Klein point, the classical criticality is dominant and deconfinement  
persists over a finite region of the phase diagram. This type of deconfined  
behavior goes well beyond the simple, thermal decoupling of a system (to  
obtain quasi--one--dimensional behavior) in that the classical critical  
point is driven to $T = 0^+$: the microscopic origin of deconfinement lies  
in the proximity of the system to a Klein point, and spinon propagation  
remains $d$--dimensional. 
 
We have obtained our conclusions, and as a result a rather  
complete picture of a distinctive paradigm for classical and  
quantum criticality, from a number of rigorous techniques. The  
dimer coverings of the checkerboard and pyrochlore lattices may be  
treated, through their connection with the ice rules, by both the  
six--vertex mapping and a line representation, which exploit the  
underlying topological order of the physical system to classify  
the states of the manifold. We have developed a loop representation  
of the physical processes in the dimer basis which has very general  
applicability to calculations involving the non--orthogonal dimer  
states; this gives a clear and intuitive picture of all contributing  
local processes, reflected in the sizes of corresponding loops, and  
a straightforward but rigorous set of rules systematizing their  
computation.  

\acknowledgments 

We thank B. Kumar
for helpful discussions. 
We also wish to thank F. Pollmann and K. Shtengel
for their enthusiastic interest and for 
encouraging us to emphasize the absence 
of the Rokhsar-Kivelson type processes
in our bare systems. This work was sponsored by  
the U.S.~DoE under Contract No.~W--7405--ENG--36 and by PICT 03--06343  
of the ANPCyT of Argentina. 

{\bf{Note added in proof}}

Two months after the initial appearance
of our work \cite{preprint}, Pollmann
and coworkers reported, in a 
series of preprints \cite{pollmann}, on a spinless 
Fermi model which possesses 
several of the qualitative
features which we found here
for spin systems on the pyrochlore 
and checkerboard lattices.


\begin{thebibliography}{99} 
 
\bibitem{Anderson87} 
P. W. Anderson, G. Baskaran, Z. Zou, and T. Hsu, Phys. Rev. Lett. {\bf 58},  
2790 (1987). 
 
\bibitem{Polyakov77} 
A. M. Polyakov, Nucl. Phys. B {\bf 120}, 429 (1977). 
 
\bibitem{Herbut03} 
S. Sachdev and K. Park, Ann. Phys. (N. Y.) {\bf 298}, 58 (2002); 
I. F. Herbut, B. H. Seradjeh, S. Sachdev, and G. Murthy, Phys. Rev. B  
{\bf 68}, 195110 (2003). 
 
\bibitem{Herm} M. Hermele, T. Senthil, M. P. A. Fisher, P. A. Lee, N. 
Nagaosa, and X--G. Wen, Phys. Rev. B {\bf 70}, 214437 (2004). 
 
\bibitem{senthil04} T. Senthil, A. Vishwanath, L. Balents, S. Sachdev,  
and M. P. A. Fisher, Science {\bf 303}, 1490 (2004), and references therein. 
 
\bibitem{BN} C. D. Batista and Z. Nussinov, Phys. Rev. B  
{\bf 72}, 045137 (2005). 
 
\bibitem{rbt} 
C. D. Batista and S. A. Trugman, Phys. Rev. Lett. {\bf 93}, 217202 (2004). 
 
\bibitem{rramirez} A. P. Ramirez, Ann. Rev. Mater. Sci. {\bf 24}, 453 (1994). 
 
\bibitem{rtakagi} H. Takagi,

http://online.itp.ucsb.edu/online/exotic$_c$04/takagi/pdf/Takagi.pdf  
(unpublished). 
 
\bibitem{fowler} D. Bernal and R. H. Fowler, J. Chem. Phys.  
{\bf 1}, 515 (1933). 
 
\bibitem{Pauling} L. Pauling, {\em The Nature of the Chemical Bond}  
(Cornell University Press, Ithaca, 1939). 
 
\bibitem{frustration} R. Moessner, Can. J. Phys. {\bf 79}, 1283 (2001)  
and references therein; R. Moessner and J. T. Chalker, Phys. Rev. Lett.  
{\bf 80}, 2929 (1998). 
 
\bibitem{Canals} B. Canals and C. Lacroix, Phys. Rev. Lett., {\bf 80},  
2933 (1998). 
 
\bibitem{gingras}S. T. Bramwell and M. J. P. Gingras, Science 
{\bf 294}, 1495 (2001), and references therein. 
 
\bibitem{conflict} 
We provide a brief summary of selected contributions to the pyrochlore  
problem. The case of an Ising antiferromagnet on a pyrochlore lattice 
was investigated by Anderson \cite{anderson}, who noted the similarity to  
the spin--ice problem. Analytical treatments of this model \cite{dipole}  
have yielded the precise form of the correlation functions, which have 
been confirmed by numerical studies \cite{numericalIsing}. The classical 
antiferromagnet has been investigated by the large--$N$ technique 
\cite{largen}, and quantum Heisenberg spins on the pyrochlore structure have 
been addressed in a number of studies \cite{ring_exchange1,ring_exchange2} 
by perturbative treatments based on the Ising result with sufficiently large 
transverse terms. These approaches lead to the rich physics of ring--exchange  
interactions \cite{ring_exchange1,ring_exchange2}, as well as to a rather 
elegant charge model \cite{shannon}, and suggest the strong possibility of 
fractionalized excitations. While some approaches \cite{Canals} indicate 
the absence of spin order in the quantum pyrochlore antiferromagnet, 
numerical studies suggest specific singlet ordering patterns for the 
ground states of certain $S = 1/2$ antiferromagnetic models \cite{Sondhi, 
canals, erez}. Specific results for simple, microscopic, isotropic, 
nearest--neighbor quantum spin Hamiltonians are lacking. Contradicting 
the support for fractionalization, the large--$S$ technique suggests for 
the classical Heisenberg antiferromagnet on the pyrochlore and checkerboard 
lattices \cite{oleg,uzi} that magnetic order may be established at low 
temperatures, driven by an order--by--disorder \cite{frustration,villain} 
effect where tunneling between states lifts the classical ground--state 
degeneracy and stabilizes order. Recent studies \cite{orbital} of systems 
similar to the effective spin models of Ref.~\cite{erez} have provided 
rigorous demonstrations of this effect for the classical case 
({\it i.e.}~with thermal rather than quantum fluctuations).
 
\bibitem{anderson} P. W. Anderson, Phys. Rev. {\bf 102}, 1008 (1956). 
 
\bibitem{dipole} J. Villain, Solid State Comm. {\bf 10}, 967 (1972); 
F. H. Stillinger and M. A. Cotter, J. Chem. Phys. {\bf 58}, 2532 (1973);  
R. W. Youngblood and J. D. Axe, Phys. Rev. B {\bf 23}, 232 (1981);  
L. B. Ioffe and A. I. Larkin, Phys. Rev. B {\bf 40}, 6941 (1989); 
D. A. Huse. W. Krauth, R. Moessner, and S. L. Sondhi, Phys. Rev. Lett.  
{\bf 91}, 167004 (2003); C. L. Henley, Phys. Rev. B {\bf 71}, 014424 (2005).  
 
\bibitem{numericalIsing} S. Yoshida, K. Nemoto, and K. Wada, 
J. Phys. Soc. Jpn. {\bf 71}, 948 (2002). 
 
\bibitem{largen} D. A. Garanin and B. Canals, Phys. Rev. B {\bf 59},  
443 (1999); B. Canals and D. A. Garanin, Can. J. Phys. {bf 79}, 1323 (2001). 
 
\bibitem{ring_exchange1} M. Hermele, M. P. A. Fisher, and L. Balents, 
Phys. Rev. B {\bf 69}, 064404 (2004) 
 
\bibitem{ring_exchange2} N. Shannon, G. Misguich, and K. Penc,  
Phys. Rev. B {\bf 69}, 220403(R) (2004). 
 
\bibitem{shannon} P. Fulde, K. Penc, and N. Shannon, Ann. Phys. (Leipzig) 
{\bf 11}, 892 (2002). 
 
\bibitem{oleg} O. Tchernyshyov, O. A. Starykh, R. Moessner, and 
A. G. Abanov, Phys. Rev. B {\bf 68}, 144422 (2003). 
 
\bibitem{uzi} U. Hizi, P. Sharma, and C. L. Henley, Phys. Rev. Lett.,  
{\bf 95}, 167203 (2005). 
 
\bibitem{Sondhi} R. Moessner, O. Tchernyshyov and S. L. Sondhi, J. Stat.  
Phys. 116, 755 (2004). 
 
\bibitem{canals} B. Canals, Phys. Rev. B {\bf 65}, 184408 (2002). 
 
\bibitem{erez} E. Berg, E. Altman, and A. Auerbach,  
       Phys. Rev. Lett. {\bf 90}, 147204 (2003). 
 
\bibitem{villain} J. Villain, R. Bidaux, J. P. Carton, and R. Conte, 
Journal de Physique {\bf 41}, 1263 (1980);  
E. F. Shender, Sov. Phys. JETP {\bf 56}, 178 (1982). 
 
\bibitem{orbital} Z. Nussinov, M. Biskup, L. Chayes, and J. van den Brink, 
Europhysics Letters {\bf 67}, 990 (2004); M. Biskup, L. Chayes, Z. Nussinov, 
Comm. Math. Phys. {\bf 255}, no. 2, 253 (2005). 
 
\bibitem{explain_frustration} 
Antiferromagnetic nearest--neighbor interactions on  
the pyrochlore lattice have a strong, intrinsic geometrical frustration.  
For both classical and quantum spins it is impossible to bring all the  
individual exchange interactions close to their minimal value simultaneously.  
The competition of individual Heisenberg interactions is exacerbated by the  
inclusion of the symmetric four--spin interaction. The extent of the  
frustration can be characterized quite generally by the ground--state 
degeneracy, and this is maximized at the point $K = K_{c} = 4J/5$. A  
discussion of the nature of the level--crossing occurring at this point  
is contained in Sec.~V. 
 
\bibitem{simple_identities} The reader may wish to verify the mapping  
between Eqs.~(\ref{esh}) and (\ref{symHam}) by application of the  
standard $S = 1/2$ identities 
\begin{eqnarray} 
\{(\vec{S}_{i} \cdot \vec{S}_{j}), (\vec{S}_{i} \cdot \vec{S}_{k})\} & = 
& \frac{1}{2} (\vec{S}_{i} \cdot \vec{S}_{k}), ~~~~ j \neq k, \nonumber 
\\ (\vec{S}_{i} \cdot \vec{S}_{j})^{2}  
& = & \frac{3}{16} - \frac{1}{2} (\vec{S}_{i} \cdot \vec{S}_{j}). 
\label{si} 
\end{eqnarray} 
The first relation follows from the anticommutation of different spin  
components at a single lattice site, 
\begin{equation} 
\{ S_{i}^{a}, S_{i}^{b} \} = \frac{1}{2} \delta_{ab}, ~~~~ 
a,b =  x,y,z, 
\end{equation} 
and the second follows, among other derivations, from the identity 
$\vec{S}_{i} \cdot \vec{S}_{j} = \frac{1}{2} (P_{ij} - \frac{1}{2})$, 
where $P_{ij}$ is the operator permuting sites $i$ and $j$. 
 
\bibitem{note0} 
In the case of the checkerboard lattice there is also a biquadratic  
spin term which involves the four sites of each uncrossed plaquette.  
These processes are not included in the Hamiltonian of Eq.~(\ref{symHam}). 
 
\bibitem{rmhr} 
E. M\"uller--Hartmann and A. Reischl, Eur. Phys. J. B {\bf 28}, 173 (2002). 
 
\bibitem{rkk} 
A. A. Katanin and A. P. Kampf, Phys. Rev. B {\bf 66}, 100403(R) (2002);  
{\bf 67} 100404(R) (2003). 
 
\bibitem{rcgb} 
A. Chubukov, E. Gagliano, and C. Balseiro, Phys. Rev. B {\bf 45}, 7889 
(1988).

\bibitem{ferrom}
This lies outside the range of validity of 
${{\tilde H}_{\rm Hubb}}$ as an effective low energy Hamiltonian for
$H_{\rm Hubb}$ ($J_1=-J_2$  for $t/U = 1/\sqrt{30}$ to ${\cal{O}}(t^{4}/U^{3})$). 
However, the Klein condition $J_1=-J_2$ can be obtained for smaller values of 
$t/U$, such that ${{\tilde H}_{\rm Hubb}}$ becomes a valid description
of the low energy spectrum, if a ferromagnetic nearest--neighbor direct 
exchange interaction is added to the Hubbard model. Such a direct
exchange cotribution is always present in real systems.

\bibitem{Klein82} 
D. J. Klein, J. Phys. A: Math. Gen. {\bf 15}, 661 (1982). 
For completeness, we remark that 
Klein's original work examined 
the total spin formed by sites which 
surround a given lattice site. This is slightly different
than the case that we examine in our work. 
Here, the projection operators act on the total
spin of all sites compromising
a basic building block such as the tetrahedron.
Such a similar extended idea was
employed in \cite{rbt}.
 
\bibitem{nussinov_June06} 
Z. Nussinov, unpublished (cond--mat/0606075).
  
\bibitem{Raman05} K. S. Raman, R. Moessner, and S. L. Sondhi, 
Phys. Rev. B {\bf 72}, 064413 (2005). 
 
\bibitem{Onsager} L. Onsager and M. Dupuis, in 
{\em Rendiconti della Scuola Internationale di Fisica (Enrico  
Fermi) X Coroso}, Societa Italiana di Fisica, Bologna (1960); 
B. Simon, {\em The Statistical Mechanics of Lattice Gases}, Volume I 
Princeton University press (1993), esp.~p.~315. 
 
\bibitem{rnagle} J. F. Nagle, J. Math. Phys. {\bf 7}, 1484, (1966). 
 
\bibitem{Baxter} R. J. Baxter, {\em Exactly solved Models in Statistical 
Mechanics} (Academic Press, London, 1982).  
 
\bibitem{Lieb} E. H. Lieb, Phys. Rev. {\bf 162}, 162 (1967).  
 
\bibitem{note2} 
Two arbitrary states are in general non--orthogonal, but their overlap  
is exponentially small in the length of closed loops obtained by  
superposing them. Linear independence of singlet dimer coverings has been  
proven for other lattices in Ref.~\cite{Chayes89}. 
 
\bibitem{Chayes89} 
J. T. Chayes, L. Chayes, and S. A. Kivelson, Commun. Math. Phys. {\bf 123},  
53 (1989). 

\bibitem{rbtfig} A schematic representation of such states on the square 
lattice can be found in Figs.~1(c,d) of \cite{rbt}; these are also allowed 
states of the ground--state manifold on the checkerboard lattice. 
 
\bibitem{note3} 
We assume here that the diagonal elements of the perturbative  
term, $H'$, do not break the degeneracy; this would be the  
case when, for example, $H'$ is a Heisenberg Hamiltonian of the type  
given by the first term of $H$.  
 
\bibitem{Wenbook} X.--G. Wen, {\em Quantum Field Theory of Many--Body  
Systems} (Oxford University Press, Oxford, 2004).  
 
\bibitem{AKLT}   
I. Affleck, T. Kennedy, E. H. Lieb, and H. Tasaki, Commun. Math. Phys. 
{\bf 115}, 477 (1988). 
 
\bibitem{KT}  
T. Kennedy and H. Tasaki,  Comm. Math. Phys. {\bf 147}, 431 (1992). 
 
\bibitem{Kasteleyn63} 
P. W. Kasteleyn, Physica {\bf 27}, 1209 (1961); J. Math. Phys. {\bf 4},  
287 (1963). 
 
\bibitem{Fisher63} 
M. E. Fisher, Phys. Rev. {\bf 124}, 1664 (1961); 
P. W. Kasteleyn, Physica {\bf 27}, 1209 (1961); 
M. E. Fisher and J. Stephenson, Phys. Rev. {\bf 132}, 1411 (1963). 
 
\bibitem{note4}  
In contrast to the mappings between quantum spin and dimer models,  
orthogonality between different dimer coverings is not a requirement  
in this case because the effective spinon--spinon interaction has an  
entropic origin and the entropy is determined solely by the number of  
linearly independent states. 
 
\bibitem{Kogut} 
J. B. Kogut, Rev. Mod. Phys. {\bf 51}, 659 (1979); E. Fradkin and S. H. 
Shenker, Phys. Rev. D {\bf 19}, 3682 (1979). 
 
\bibitem{CF} 
A. H. Castro Neto, P. Pujol, and E. Fradkin, unpublished (cond-mat/0511092). 
 
\bibitem{Krauth03} 
W. Krauth and R. Moessner, Phys. Rev. B {\bf 67}, 064503 (2003). 

\bibitem{012_note}
The energy of Eq.~(\ref{symHam}) when evaluated for a single tetrahedron 
at $K = K_{c} + \delta K$ is
\begin{eqnarray}
E_{S_{\boxtimes}=0} = 0, \nonumber
\\ E_{S_{\boxtimes}=1} = \frac{5}{4} \delta K, \nonumber
\\ E_{S_{\boxtimes}=2} = \frac{12}{5} J - \frac{3}{4} \delta K,
\end{eqnarray}
for the singlet, triplet, and quintet states respectively. For 
$|\delta K| \ll  J$, the energy difference between the singlet and 
triplet states is very much smaller than the gap to the quintet state.

\bibitem{Peierls} R. E. Peierls, Phys. Rev. {\bf 54}, 918 (1938). 
 
\bibitem{Kleinert} H. Kleinert, {\em Gauge Fields in Condensed Matter  
Physics} (World Scientific, Singapore, 1989). 
 
\bibitem{XY_duality} J. Villain, Journal de Physique {\bf 36}, 581 (1975). 
 
\bibitem{elastic_duality} J. Zaanen, Z. Nussinov, and S. Mukhin,  
Ann. Phys. {\bf 310}, 181 (2004); V. Cvetkovic, Z. Nussinov, and J. Zaanen,  
to appear in Phil. Mag. B (cond--mat/0508664).   
 
\bibitem{Eduardo} E. Fradkin, {\em Field Theories of Condensed Matter  
Systems} (Addison--Wesley, Boston, 1994), and references therein. 
 
\bibitem{earlierRVB} 
G. Rumer, G\"ottinger Nachr. 337 (1932); S. Liang, B.  
Doucot, and P. W. Anderson, Phys. Rev. Lett. {\bf 61}, 365 (1988); 
B. Sutherland, Phys. Rev. B {\bf 37}, 3786 (1988); M. Kohmoto and Y.  
Shapir, Phys. Rev. B {\bf 37}, 9439 (1988);  R. McWeeny, {\em Valence  
Bond Theory and Chemical Structure (Studies in Physical and Theoretical  
Chemistry, vol.~64)}, eds. D. J. Klein and N. Trinajstic (Elsevier,  
Amsterdam, 1990) p.~13; M. Havilio, Phys. Rev. B {\bf 54}, 11929 (1996). 
 
\bibitem{RK} S. A. Kivelson, D. S. Rokhsar, and J. P. Sethna, Phys. 
Rev. B {\bf 35}, 8865 (1987); D. S. Rokhsar and S. A. Kivelson,  
Phys. Rev. Lett. {\bf 61}, 2376 (1988). 

\bibitem{kevin} 
A separate algorithm for the computation of energies and correlation 
functions in valence--bond states was developed shortly after this work 
by K. S. D. Beach and A. W. Sandvik, Nucl. Phys. B {\bf 750}, 142 (2006). 
The approach of these authors does not apply the topological rules 
exploited here, but offers insight into a number of other related points.  

\bibitem{Horn} 
D. Horn, Phys. Lett. B {\bf 100}, 149 (1981). 
 
\bibitem{Moessner2000} 
R. Moessner and S. L. Sondhi, Phys. Rev. B {\bf 63}, 224401 (2001). 

\bibitem{FMO}  S. Furukawa, G. Misguich, and M. Oshikawa,
Phys. Rev. Lett. {\bf 96},  047211 (2006).

\bibitem{explain_Young} An introductory exposition to 
Young tableaux is available in many texts (e.g. \cite{Sakurai}, 
\cite{wkt}).
Each of the Klein model ground states
is a superposition of a product
of singlets\cite{nussinov_June06}. In the Young tableaux
notation, these ground states therefore lie
exclusively in the two row
sector in which the length
of both rows is the same. The
top row labels in the Young tableaux 
denotes sites in which up
spins are antisymmetrized (to form 
singlet dimers) with
down state spins of the belonging 
to site indices in the row
just below it.

Consider now any of these ground
states ($\{|\psi_{j} \rangle\}$). In order
for an operator $V$ acting on these
states ($V | \psi_{j} \rangle$) to lie
in the same Young tableaux sector
and allow for finite matrix
elements $\langle \psi_{i} | V | \psi_{j} \rangle$
in the Klein ground state basis,
we must have that $V$ merely permutes the
site indices on the top and bottom
rows. The operator $V$ can, of course, also
do nothing at all (and act as the identity);
this is the case for a uniform rotation
operator which leaves the Klein spin state
invariant. [After all, any Klein basis
spin state is a global singlet ($S_{tot} =0$)
as it is a superposition of a product of singlet dimers.]
Thus, the most general operator
$V$ which has non-vanishing matrix
elements in the Klein ground state
basis has components which are
the a superposition of the most
general permutation operations
(including the trivial
permutation- the identity
operator).  

We re-emphasize that this follows
as (i) the different Young tableaux
sectors are orthogonal to
one another and that (ii) within 
the Young tableaux pertinent 
to the Klein basis sector, all
sites on the top row of the tableaux
lie in the up spin state
while all sites in the bottom
row (whose length is equal 
to that of the top row)  
are indexed by a
down spin state. No 
operators $V$ other than 
those involving a sum of permutations
can link states in this 
Young tableaux sector.


\bibitem{Sakurai}
J. J. Sakurai, {\em Modern Quantum Mechanics}
(Addison-Wesley Publishing Company (1985)), 
[see chapter 6 in particular].

\bibitem{wkt} Wu-Ki Tung, {\em Group Theory in Physics} 
(World Scientific Publishing Company (1985)) 
[see chapters 5 and 13 in particular]. 

\bibitem{entropy-energy} 
In the absence of external symmetry--breaking fields, the competition 
of energy and entropy is resolved universally in favor of disorder at 
sufficiently high temperatures (symmetry restoration). We summarize 
those parts of the physical origin of this phenomemon most relevant 
to the current considerations. General disordering events (similar to 
those in Sec.~IV) have positive energy penalties ($\Delta E$) and entropy 
gains ($\Delta S$). At sufficiently high temperatures $T$, the free--energy 
change due to the increase in disordering processes, $\Delta F = \Delta E - 
T \Delta S$, becomes negative: defects proliferate and order is destroyed. 
For general, discrete order parameters in a system of finite--range 
interactions, the scaling of $\Delta E$ with defect (domain--wall) size 
is no faster than that of $\Delta S$, whence domain--wall deconfinement 
leads to a lower free energy at sufficiently high $T$. For continuous 
order parameters there exist further, lower--energy, deformations in 
addition to domain walls, which are already favorable. Thus there is 
always a temperature $T_{c}$ beyond which the free energy is lowered 
monotonically
by contributions from disorder--increasing processes. As a consequence, 
for $T > T_{c}$ the expectation value $\langle \hat{O}_{\bf r} \rangle = 0$. 

\bibitem{exception} The development of long--range order in $Q$ [$m \equiv 
\langle Q_{A} \rangle \neq 0$ at all finite $T$] in this case is not in 
conflict with finite--temperature singularities in the free energy which 
may lead to algebraic correlations in quantities other than $Q$. One example 
is afforded by the classical six--vertex model in any symmetry--breaking 
field, which is analogous to the parameter $\epsilon$ in $H'$. In this 
model the field may cause all six states to have different energies, 
but criticality appears nevertheless at finite fields and temperatures 
[see for example Fig.~8.6 in Ref.~\cite{Baxter}]. 

\bibitem{marginalconfinement} 
We should remark that this finite $T$ critical behavior is distinct 
from the exact deconfinement present at $T=0$ albeit very
similar to it. At $T>0$, the entropic contribution $(-T \Delta S)$ to 
the free energy leads to an effective (marginally confining) 
logarithmic interaction between monomers at the ice point. However, 
as discussed in Sec.(\ref{thermalklein}), this logarithmic interaction 
is screened (by virtue of a finite defect concentration at $T>0$) 
at sufficiently large distances. 

\bibitem{Batista04} 
C. D. Batista and G. Ortiz, Adv. Phys. {\bf 53}, 1 (2004). 
 
\bibitem{rcastelnovo} C. Castelnovo, C. Chamon, C. Mudry, and P. Pujol,  
Phys. Rev. B {\bf 73}, 144411 (2006). 
 
\bibitem{kirk} T. R. Kirkpatrick and P. G. Wolynes, Phys. Rev. B {\bf 
35}, 3072 (1987); {\bf 36}, 8552 (1987); T. R. Kirkpatrick and 
P. G. Wolynes, Phys. Rev. A {\bf 40}, 1045 (1989); T. R. Kirkpatrick 
and D. Thirumalai, Phys. Rev. Lett. {\bf 58}, 2091 (1987). 
 
\bibitem{Kauzmann} W. Kauzmann, Chem. Rev. {\bf 43}, 219 (1948). 
 
\bibitem{Chamon} 
C. Chamon, Phys. Rev. Lett. {\bf 94}, 040402 (2005); C. Castelnovo,  
C. Chamon, C. Mudry, and P. Pujol, Phys. Rev. B {\bf 72}, 104405 (2005). 
 
\bibitem{coleman} P. Chandra, P. Coleman, and I. Ritchie,  
J. Phys. I (France) {\bf 3}, 591 (1993). 
 
\bibitem{Gardner} J. S. Gardner, G. Ehlers, S. T. Bramwell, and B. D. Gaulin,  
J. Phys: Condens. Matter {\bf 16}, S643 (2004). 
 
\bibitem{Ladieu} F. Ladieu, F. Bert, V. Dupuis, E. Vincent, and J. Hammann, 
J. Phys: Condens. Matter {\bf 16}, S735 (2004). 
 
\bibitem{Bono} D. Bono, P. Mendels, G. Collin, N. Blanchard, C. Baines, and  
A. Amoto, J. Phys: Condens. Matter {\bf 16}, S817 (2004). 
 
\bibitem{Nakatsuji} 
S. Nakatsuji, Y. Nambu, H. Tonomura, O. Sakai, S. Jonas, C. Broholm,  
H. Tsunetsugu, Y. Qiu, and Y. Maeno, Science {\bf 309}, 1697 (2005). 

\bibitem{preprint}  Z. Nussinov, C. D. Batista, B. Normand, S. A. Trugman,
cond-mat/0602528  (2006).

\bibitem{pollmann}  F. Pollmann, P. Fulde, E. Runge,  Phys. Rev. B 73, 125121 
(2006) [cond-mat/0604122];  F. Pollmann, P. Fulde, Europhys. Lett., 75 (1), 
pp. 133-138 (2006) [cond-mat/0604666]; 
 F. Pollmann, J. J. Betouras, K. Shtengel, P. Fulde, cond-mat/0607202 (2006);
F. Pollmann, J. J. Betouras, E. Runge, P. Fulde, cond-mat/0609122 (2006)
\end{thebibliography}
\end{document}